\documentclass[fleqn,usenatbib]{mnras}

\usepackage{newtxtext,newtxmath}

\usepackage[T1]{fontenc}

\DeclareRobustCommand{\VAN}[3]{#2}
\let\VANthebibliography\thebibliography
\def\thebibliography{\DeclareRobustCommand{\VAN}[3]{##3}\VANthebibliography}

\usepackage{graphicx}	
\usepackage{amsmath}	

\usepackage[dvipsnames]{xcolor}

\newcommand{\rev}[1]{#1}

\title[Age-dependent Galactic metallicity gradient]{Age dependence of the radial metallicity gradient in the Galactic disc with LAMOST and Gaia}

\author[]{
Yi-De Jing,$^{1,2}$\thanks{E-mail: \href{mailto:jingyide@qhu.edu.cn}{jingyide@qhu.edu.cn}} 
Duojie Jia,$^{3,4}$\thanks{E-mail: \href{mailto:jiadj@nwnu.edu.cn}{jiadj@nwnu.edu.cn}}
Bin Wang,$^{1,2}$
Yulong Gao,$^{1,2}$
Liyuan Lu$^{1,2}$
Zheng-Jun Du$^{5,6}$
and Zenghui Zou$^{1,2}$
\\
$^{1}$Department of Physics and Astronomy, Qinghai University, Xining, 810016, China\\
$^{2}$Research Center of Astronomy, Qinghai University, Xining, 810016, China\\
$^{3}$General Education Center, Qinghai Institute of Technology, No. 2, Xining, 810016, China\\
$^{4}$Lanzhou Center for Theoretical Physics, Key Laboratory of Theoretical Physics of Gansu Province, Key Laboratory of Quantum Theory and Applications of MoE, \\ Gansu Provincial Research Center for Basic Disciplines of Quantum Physics, Lanzhou University, Lanzhou 730000, China\\
$^{5}$School of Computer Technology and Application, Qinghai University, Xining 810016, China\\
$^{6}$Qinghai Provincial Key Laboratory of Media Integration Technology and Communication, Qinghai University, Xining 810016, China
}

\date{Accepted XXX. Received YYY; in original form ZZZ}

\pubyear{2026}

\begin{document}
\label{firstpage}
\pagerange{\pageref{firstpage}--\pageref{lastpage}}
\maketitle


\begin{abstract}

We investigate the \rev{stellar radial metallicity ([Fe/H])} gradient--age relation (MGAR) of the Galactic disc using a solar-neighbourhood sample of 580\,037 FGK dwarf and turn-off stars from LAMOST DR7, Gaia DR3, and the corrected StarHorse age catalogue. The fiducial MGAR is non-monotonic, with relatively strong negative gradients at intermediate ages followed by substantial flattening towards older populations. The most negative gradient occurs at $6$--$8$ Gyr in the adopted 2-Gyr bins, although the $4$--$6$ Gyr slope is statistically similar, indicating a broad intermediate-age turnover rather than a sharply defined transition. The flattening persists within the low-$\alpha$ population, indicating that chemical-population mixing alone cannot explain the full-sample relation, and is also recovered under alternative guiding-radius definitions, radial fitting ranges, vertical-orbit selections, propagation of the reported random age uncertainties, \rev{and an alternative stellar-age scale.} In contrast, controlling the age-dependent $T_{\rm eff}$--$\log g$ distribution substantially reduces the absolute gradient amplitudes and modifies the younger-age behaviour. \rev{A same-star comparison with APOGEE DR17 further shows that the broad age dependence of the controlled MGAR is recovered using an external [Fe/H] scale.} We therefore identify the intermediate-to-old-age flattening as the most robust feature of the observed MGAR, whereas the precise location and depth of the intermediate-age minimum and the detailed younger-age morphology are less stable. These results indicate that stellar-parameter selection can contribute to the diversity of MGAR morphologies reported in previous studies.

\end{abstract}



\section{Introduction}

The Milky Way provides a unique nearby laboratory for studying the formation and evolution of disc galaxies. Its present-day stellar populations retain information about the chemical enrichment and dynamical evolution of the Galactic disc, allowing its history to be investigated through stellar ages, chemical abundances, and orbital properties \citep{Freeman2002,BlandHawthorn2016,Matteucci2026}. Large spectroscopic surveys, including APOGEE \citep{Majewski2017}, GALAH \citep{DeSilva2015}, LAMOST \citep{Zhao2012,Deng2012}, SDSS-V \citep{Kollmeier2026}, and the DESI Milky Way Survey \citep{DESIMWS2022}, together with Gaia astrometry \citep{Gaia2016,Gaia2023}, now make it possible to examine these relations for unprecedented numbers of stars. Stellar age is particularly important because it provides the temporal information required to connect present-day Galactic structure with its evolutionary history \citep{Casagrande2011,Miglio2013}.

Among the resulting observables, the stellar radial metallicity gradient provides an empirical measure of the spatial variation of chemical enrichment across the disc. \rev{Throughout this work, the stellar metallicity gradient refers specifically to the radial gradient in [Fe/H]. This distinction is important when comparing stellar and interstellar-medium abundance gradients: iron is widely used as a metallicity tracer in stellar studies, whereas the metallicity of ionized gas is commonly characterised using O/H. Because Fe and O have different nucleosynthetic origins and enrichment timescales, their radial abundance gradients are not directly interchangeable \citep{MendezDelgado2024}.} Negative radial gradients arise naturally in inside-out disc-evolution models in which the inner disc experiences more rapid star formation and enrichment \citep{Matteucci1989,Chiappini1997,Matteucci2001,Minchev2013}, and are well established observationally using a wide variety of stellar tracers \citep{Genovali2014,Bergemann2014,Netopil2016,Anders2017,Queiroz2023}.

A key unresolved question is how the radial metallicity gradient depends on stellar age. Early studies of nearby F- and G-dwarf samples already indicated that the radial metallicity gradient varies among stellar populations of different ages \citep{Nordstrom2004,Casagrande2011}. Measurements based on larger spectroscopic samples, however, do not converge on a single morphology of the metallicity gradient--age relation (MGAR). Studies based on main-sequence turn-off and subgiant populations have reported flatter gradients among the youngest and oldest stars, with stronger negative gradients at intermediate ages \citep{Xiang2015,Wang2019}. Non-monotonic behaviour has also been reported in some red-giant samples, although the location and strength of the steepest gradient differ from those inferred for MSTO stars \citep{Anders2017,Anders2023}. In contrast, other analyses based on guiding radii or giant and red-clump populations favour a smoother flattening of the gradient towards older ages \citep{Vickers2021,Willett2023,Chen2026,Sun2026}. Recent comparisons across stellar tracers therefore highlight substantial differences in the inferred MGAR morphology \citep{Wang2023a,Chen2026}. The existence, location, and amplitude of an intermediate-age minimum consequently remain uncertain, and it is not yet clear to what extent the diversity of published MGARs reflects genuine Galactic evolution or differences in stellar tracers, radial coordinates, age scales, abundance measurements, and sample selection.

The physical interpretation of the observed MGAR is further complicated because it is not a direct record of the temporal evolution of the gas-phase abundance gradient. \rev{In the ionized interstellar medium, the present-day Galactic metallicity gradient is commonly characterised using O/H measurements of H\,{\sc ii} regions \citep{MendezDelgado2022}, while comparisons between H\,{\sc ii} regions and planetary nebulae provide complementary constraints on the temporal evolution of nebular abundance gradients \citep{Amayo2026}. These O/H-based abundance gradients should not be directly identified with the stellar [Fe/H] gradients studied here because Fe and O trace different nucleosynthetic channels and enrichment timescales \citep{MendezDelgado2024}.}

A present-day mono-age stellar population reflects both the radial chemical structure of the interstellar medium from which it formed and the subsequent evolution of its orbital distribution. Churning can change stellar angular momentum and guiding-centre radius, whereas orbital blurring changes the instantaneous radius sampled along an orbit without necessarily altering the guiding centre \citep{Sellwood2002,Schonrich2009,Minchev2013,Frankel2018,Frankel2020}. Consequently, neither the present-day Galactocentric radius nor the present-day guiding radius can generally be identified with the stellar birth radius. Reconstructions of stellar birth radii and birth metallicity gradients likewise show that the observed MGAR and the temporal evolution of the birth gradient need not be identical \citep{Minchev2018,Ratcliffe2023,Lu2024}. 
Moreover, chemical-evolution models predict that the \rev{gas-phase abundance gradient} can itself evolve substantially with time \citep{Sharda2021a}, while recent cosmological simulations indicate that part of the present-day age dependence of stellar metallicity gradients may already be imprinted at birth rather than being generated entirely by subsequent radial redistribution \citep{Graf2025a}. The observed MGAR must therefore be interpreted in the context of both time-dependent chemical enrichment and subsequent stellar dynamical evolution.

A further complication is the changing mixture of Galactic disc populations with stellar age. The metallicity--$\alpha$-abundance plane contains prominent low-$\alpha$ and high-$\alpha$ sequences with distinct chemical and spatial properties \citep{Fuhrmann1998,Adibekyan2012,Hayden2015,Jing2026}. High-$\alpha$ stars are preferentially old, whereas younger populations are dominated by the low-$\alpha$ sequence, although substantial overlap remains at intermediate ages \citep{Haywood2013,Feuillet2019,Queiroz2023,Xiang2022}.  The radial metallicity structure also differs between chemically and vertically distinct populations, with low-$\alpha$ populations generally exhibiting stronger radial gradients than more $\alpha$-enhanced populations \citep{Hayden2014,Vickers2021,Imig2023}. Moreover, chemically defined low- and high-$\alpha$ populations do not map uniquely onto geometrically defined thin and thick discs \citep{Hayden2017,Feuillet2019,Queiroz2023}. The apparent age dependence of the radial metallicity gradient may therefore be affected by systematic changes in the relative contributions of chemically and vertically distinct populations, motivating measurements of the MGAR within more homogeneous population selections.

Differences in the stellar tracers themselves may introduce an additional source of variation. In a dwarf- and turn-off-dominated sample, stellar age is coupled to position in the $T_{\rm eff}$--$\log g$ plane, so different age intervals need not contain equivalent stellar populations even when identical global selection criteria are applied. This issue is particularly relevant when comparing MGAR measurements based on MSTO, dwarf, RGB, and RC stars. The connection between effective temperature, stellar age, and the inferred radial metallicity structure has long been recognised in studies of nearby main-sequence populations \citep{Yu2012}, implying that changes in the stellar-parameter distribution can contribute to apparent differences between age-selected samples. At the same time, older Galactic disc populations are dynamically hotter and occupy broader orbital distributions, as reflected in the well-established increase of stellar velocity dispersion with age \citep{Nordstrom2004,Aumer2009,Aumer2016,YuLiu2018,Mackereth2019,Sharma2021}. These chemical, stellar-parameter, and dynamical correlations all vary with age and must be distinguished from evidence for a specific physical mechanism. In particular, present-day orbital excursions and guiding radii do not provide direct measurements of changes in angular momentum associated with churning, and correlations between age, chemistry, and kinematics alone cannot uniquely identify the dynamical process responsible for the observed age dependence of the radial metallicity gradient \citep{Sellwood2002,Schonrich2009,Minchev2013}.

The central question is therefore not simply whether the Galactic MGAR is monotonic or non-monotonic, but which features of the observed relation remain stable when the stellar populations and analysis choices used to measure it are controlled. In this work, we address this question using a large solar-neighbourhood sample of FGK dwarf and turn-off stars constructed from LAMOST DR7 spectroscopy, Gaia DR3 astrometry, and corrected StarHorse ages. We first establish the fiducial age dependence of the radial metallicity gradient and then examine its sensitivity to chemical-population definition, radial-coordinate choice, radial fitting range, vertical orbital structure, \rev{stellar-age scale,} and the propagation of the reported random stellar-age uncertainties. Particular attention is given to the age-dependent $T_{\rm eff}$--$\log g$ distribution through overlapping-domain restriction, reweighting, and hard-matching tests. \rev{We additionally use a same-star comparison with APOGEE DR17 to assess the sensitivity of the controlled MGAR to the adopted spectroscopic [Fe/H] scale.} Stellar velocity dispersions and present-day radial orbital excursions are subsequently used to provide dynamical context without assigning a causal origin to the observed age dependence of the radial metallicity gradient. By separating selection-sensitive features from those that persist under these controls, we aim to identify the most robust empirical characteristics of the Galactic MGAR and to clarify their relation to the apparently discrepant gradient--age relations reported in previous studies.

\section{Data and sample selection}\label{Sec:Data_Sample}

\subsection{Spectroscopic, astrometric, and age data}\label{Sec:Data_Age}

The spectroscopic data were taken from the LAMOST DR7 v2.0 low-resolution stellar catalogue \citep{Zhao2012,Cui2012,Deng2012}. For each star, we adopted the effective temperature ($T_{\rm eff}$), surface gravity ($\log g$), metallicity ([Fe/H]), and radial velocity from the LAMOST stellar-parameter catalogue, together with their reported uncertainties. The [$\alpha$/M] abundance was taken from the corresponding LAMOST DR7 abundance estimates. For stars with multiple LAMOST observations, we retained the spectrum with the highest $r$-band signal-to-noise ratio.

Astrometric and photometric information was taken from Gaia DR3 \citep{Gaia2016,Gaia2023}. We used the sky positions, proper motions, Gaia photometry, and astrometric-quality information required for the sample selection and the phase-space transformations described in Section~\ref{Sec:Methods}.  Stellar distances and ages were adopted from the corrected StarHorse catalogue associated with LAMOST DR7 \citep{Queiroz2018,Queiroz2023}. StarHorse combines spectroscopic information, multi-band photometry, and Gaia astrometry within a Bayesian framework to infer stellar parameters, distances, and ages. \rev{For the spectroscopic catalogues, the metallicity input to StarHorse is [M/H], obtained from the spectroscopic [Fe/H] and, when available, [$\alpha$/Fe] using the prescription of \citet{Salaris1993}; when no [$\alpha$/Fe] measurement is available, [M/H] is approximated by [Fe/H] \citep{Queiroz2023}. The LAMOST [Fe/H] measurements used in our MGAR analysis are therefore related to the metallicity information entering the StarHorse age inference, and the two quantities cannot be assumed to be statistically independent.}

For each star, the posterior median estimates \texttt{Dist50} and \texttt{Age50} were adopted as the fiducial distance and age, respectively. Their uncertainties were characterised using the corresponding 16th and 84th percentiles, with $\sigma_{\rm Dist}=({\rm Dist84}-{\rm Dist16})/2$ and $\sigma_{\rm Age}=({\rm Age84}-{\rm Age16})/2$. The asymmetric age intervals themselves were retained for the Monte Carlo uncertainty propagation described in Section~\ref{Sec:Methods}. The LAMOST, Gaia, and StarHorse catalogues were associated through the Gaia source identifier. For duplicate StarHorse entries associated with the same Gaia source, we retained the entry with the smallest relative age uncertainty. The resulting catalogue combines spectroscopy and chemical abundances from LAMOST, astrometry and photometry from Gaia, and distance and age estimates from StarHorse.

\rev{To assess the sensitivity of the controlled MGAR to the adopted spectroscopic [Fe/H] scale, we additionally used [Fe/H] measurements from APOGEE DR17 \citep{Abdurrouf2022}. APOGEE stars were cross-matched to the LAMOST sample, and the resulting common-star sample was used only for this external abundance-scale comparison; the fiducial analysis remains based on LAMOST [Fe/H].}

\subsection{Sample selection}

To construct a relatively homogeneous tracer population, we restricted the analysis to dwarf- and turn-off-dominated FGK stars with $5000<T_{\rm eff}<6800$~K and $3.6<\log g<4.5$. We further applied quality cuts on the stellar ages and distances, LAMOST signal-to-noise ratios, chemical-abundance uncertainties, radial-velocity uncertainty, and Gaia astrometric quality. In particular, the relative age and distance uncertainties were required to be below 30 per cent, together with $\sigma_{\rm [Fe/H]}<0.15$ dex, $\sigma_{[\alpha/{\rm M}]}<0.08$ dex, $\sigma_{\rm RV}<10~{\rm km~s^{-1}}$, and ${\rm RUWE}<1.4$ \citep{Lindegren2021}.

The sample was restricted to ${\rm Dist50}<2$ kpc, and additional colour--magnitude cuts were applied to retain the intended dwarf and turn-off population. We used the observed Gaia $G_{\rm BP}-G_{\rm RP}$ colour and a distance-modulus-based $G$-band absolute magnitude, $M_G=G-5\log_{10}(d/{\rm pc})+5$, computed using the StarHorse median distance. No explicit extinction correction was applied to the Gaia colour or $G$-band magnitude used for these selection cuts. After transforming the selected stars to Galactocentric coordinates, we further required $|Z|<1$ kpc, $100<V_\phi<300~{\rm km~s^{-1}}$, $V_{\rm tot,LSR}<180~{\rm km~s^{-1}}$, and $3<R_{\rm guide}<13$ kpc to reduce contamination from halo and dynamically extreme stars. The complete set of selection criteria is summarised in Table~\ref{tab:sample_selection}.

The initial LAMOST--Gaia--StarHorse cross-match contains 994\,866 stars. After applying the spectroscopic, age, distance, stellar-parameter, abundance, astrometric, and colour--magnitude selection criteria, 633\,890 stars remain. The subsequent spatial and kinematic selections yield a final sample of 580\,037 dwarf- and turn-off-dominated FGK stars.

\begin{table}
	\centering
	\caption{Summary of the adopted sample-selection criteria and sample statistics.}
	\label{tab:sample_selection}
	\renewcommand{\arraystretch}{1.08}
	\begin{tabular}{@{}p{0.52\columnwidth}l@{}}
		\hline
		\textbf{Parameter} & \textbf{Criterion} \\
		\hline
		Distance & $0<{\rm Dist50}<2.0$ kpc \\
		Relative distance error & $\sigma_{\rm Dist}/{\rm Dist50}<0.30$ \\
		Age & $0.8<{\rm Age50}<13.8$ Gyr \\
		Relative age error & $\sigma_{\rm Age}/{\rm Age50}<0.30$ \\
		${\rm SNR}_g$ & $>20$ \\
		${\rm SNR}_r$ & $>30$ \\
		$T_{\rm eff}$ & $5000<T_{\rm eff}<6800$ K \\
		$\log g$ & $3.6<\log g<4.5$ \\
		$\sigma_{\rm [Fe/H]}$ & $<0.15$ dex \\
		$\sigma_{[\alpha/{\rm M}]}$ & $<0.08$ dex \\
		$\sigma_{\rm RV}$ & $<10~{\rm km~s^{-1}}$ \\
		$[\mathrm{Fe/H}]$ & $-1.5<[\mathrm{Fe/H}]<0.6$ \\
		$[\alpha/{\rm M}]$ & $-0.15<[\alpha/{\rm M}]<0.45$ \\
		RUWE & $<1.4$ \\
		$G_{\rm BP}-G_{\rm RP}$ & $0.3<G_{\rm BP}-G_{\rm RP}<1.4$ \\
		$M_G$ & $2.5<M_G<6.5$ \\
		$V_\phi$ & $100<V_\phi<300~{\rm km~s^{-1}}$ \\
		$V_{\rm tot,LSR}$ & $<180~{\rm km~s^{-1}}$ \\
		$|Z|$ & $<1$ kpc \\
		$R_{\rm guide}$ & $3<R_{\rm guide}<13$ kpc \\
		\hline
		LAMOST--Gaia--StarHorse matched & 994\,866 \\
		Pre-kinematic selected sample & 633\,890 \\
		Final dwarf/turn-off FGK sample & 580\,037 \\
		\hline
	\end{tabular}
\end{table}

\begin{figure*}
	\centering
	\includegraphics[width=0.9\textwidth]{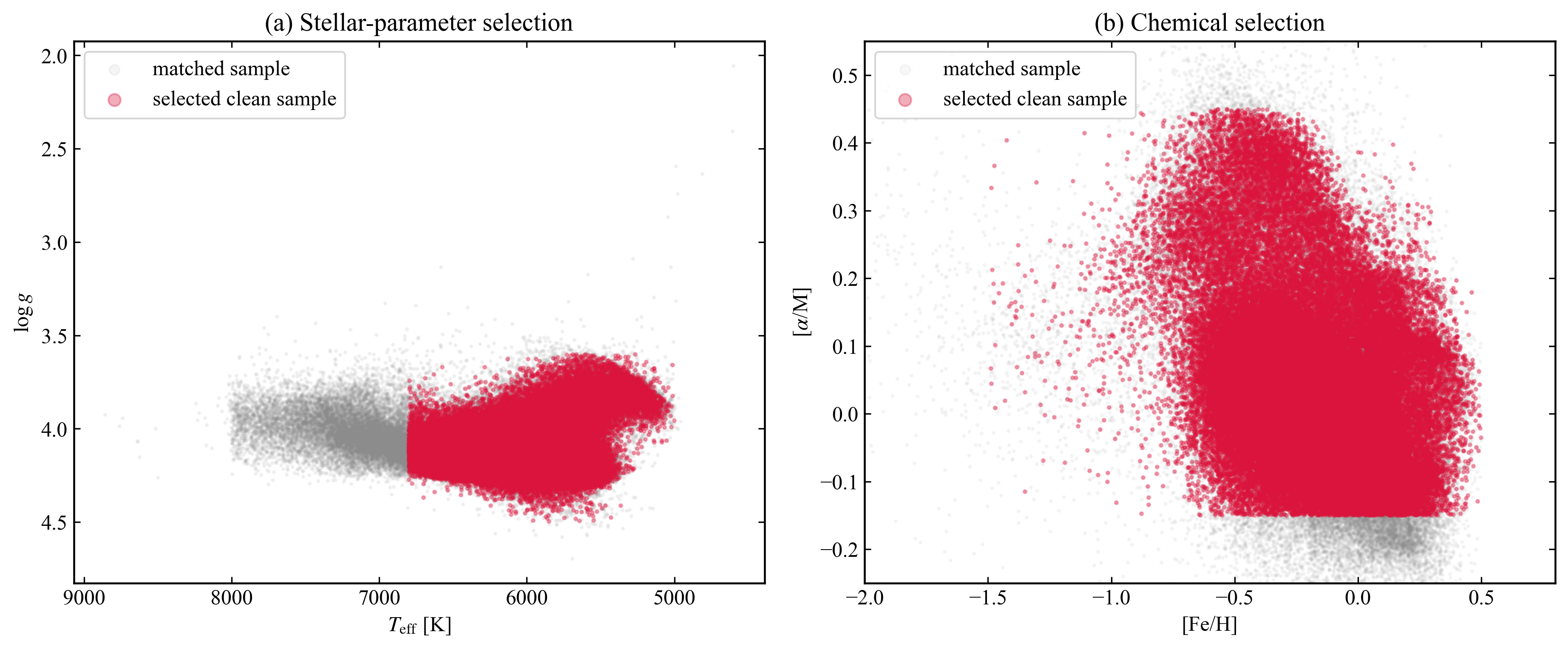}
	\caption{Selection of the final stellar sample. The left panel shows the $T_{\rm eff}$--$\log g$ distribution, and the right panel shows the [Fe/H]--[$\alpha$/M] distribution. Grey points denote the initial LAMOST--Gaia--StarHorse matched sample, while red points denote the final selected sample.}
	\label{fig:all_vs_selected_sample}
\end{figure*}

Fig.~\ref{fig:all_vs_selected_sample} illustrates the effect of the adopted selection criteria on the stellar-parameter and chemical distributions. The final sample occupies the intended dwarf and turn-off region in the $T_{\rm eff}$--$\log g$ plane while retaining the broad chemical structure of the Galactic disc.

\begin{figure*}
	\centering
	\includegraphics[width=0.9\textwidth]{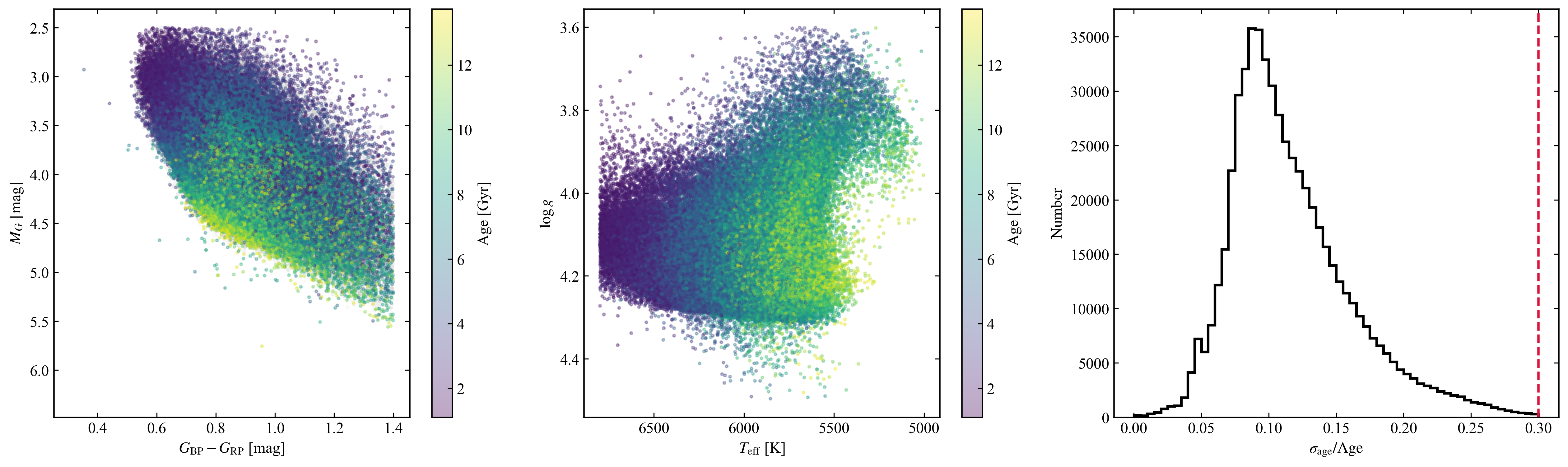}
	\caption{Distribution of the adopted stellar ages. The left panel shows the Gaia colour--magnitude diagram coloured by stellar age, and the middle panel shows the $T_{\rm eff}$--$\log g$ plane coloured by age. The right panel shows the relative age-uncertainty distribution; the red dashed line marks the adopted limit of $\sigma_{\rm Age}/{\rm Age50}=0.30$.}
	\label{fig:CMD_and_age_validation}
\end{figure*}

Fig.~\ref{fig:CMD_and_age_validation} illustrates the distribution of the adopted stellar ages across the colour--magnitude and $T_{\rm eff}$--$\log g$ planes. The systematic variation of age with stellar evolutionary position is evident, while the relative age uncertainties remain below the adopted 30 per cent threshold by construction.

\section{Analysis methods}
\label{Sec:Methods}

\subsection{Kinematics and orbital quantities}

Galactic kinematics were derived by combining Gaia DR3 astrometry with LAMOST radial velocities. The transformation to a right-handed Galactocentric cylindrical coordinate system was performed with \texttt{astropy.coordinates} \citep{Astropy2013,Astropy2018}. We adopted $R_0=8.2$ kpc \citep{Gravity2019}, $V_{\rm circ}=248$ km s$^{-1}$ \citep{Bobylev2010}, $Z_\odot=20.8$ pc \citep{Bennett2019}, and a solar peculiar motion of $(U_\odot,V_\odot,W_\odot)=(11.1,12.24,7.25)$ km s$^{-1}$ \citep{Schonrich2010}.

From the Galactocentric phase-space coordinates, we calculated the cylindrical velocity components $(V_R,V_\phi,V_Z)$ and the \rev{vertical component of the angular momentum}, $L_Z=R_{\rm GC}V_\phi$, where $R_{\rm GC}$ is the instantaneous Galactocentric cylindrical radius. Assuming an approximately flat rotation curve, we estimated the fiducial guiding radius as
\begin{equation}
	R_{\rm guide}=\frac{L_Z}{V_{\rm circ}} .
	\label{eq:Rguide}
\end{equation}
We adopted $R_{\rm guide}$ as the fiducial radial coordinate because it is less sensitive than $R_{\rm GC}$ to the instantaneous epicyclic phase \citep{Schonrich2009,Minchev2013}. As robustness tests, guiding radii and maximum orbital heights, $Z_{\max}$, were also calculated in MWPotential2014 using \texttt{galpy} \citep{Bovy2015}, scaled to the same adopted $R_0$ and $V_{\rm circ}$ as in the fiducial analysis; $Z_{\max}$ was estimated with the analytic St\"ackel approximation. These comparisons are presented in Appendix~\ref{Appendix:gradient_validation}.

The radial, azimuthal, and vertical velocity dispersions, $\sigma_R$, $\sigma_\phi$, and $\sigma_Z$, were estimated as half of the 16th--84th percentile range of the corresponding velocity distributions, with uncertainties evaluated by bootstrap resampling. We additionally used $|R_{\rm guide}-R_{\rm GC}|$ as a proxy for present-day radial orbital excursions. This quantity primarily reflects the instantaneous epicyclic displacement associated with orbital blurring, rather than changes in guiding radius or angular momentum associated with churning.

\subsection{Metallicity-gradient measurements}

For visualising the chemical distributions, we used four broad age intervals: $0.8$--$3$, $3$--$6$, $6$--$9$, and $9$--$13.8$ Gyr. The quantitative MGAR analysis instead used the non-overlapping intervals $0.8$--$2$, $2$--$4$, $4$--$6$, $6$--$8$, $8$--$10$, $10$--$12$, and $12$--$13.8$ Gyr.

The fiducial radial metallicity gradients were measured directly from individual stars over $5.0<R_{\rm guide}<10.5$ kpc. In each age bin, we performed an unweighted ordinary least-squares fit to the relation ${\rm [Fe/H]}=aR_{\rm guide}+b$, with the slope $a={\rm d[Fe/H]}/{\rm d}R_{\rm guide}$ defining the \rev{stellar} radial metallicity gradient in units of dex kpc$^{-1}$. Gradient uncertainties were estimated from 1000 bootstrap realisations obtained by resampling stars with replacement within each age interval.

As a complementary check, we also fitted median [Fe/H] profiles constructed in 0.5-kpc $R_{\rm guide}$ bins. These median-profile gradients were used only as a robustness diagnostic. Additional tests involving radial fitting intervals, guiding-radius definitions, vertical-orbit selections, and propagation of the reported random stellar-age uncertainties are presented in Appendix~\ref{Appendix:gradient_validation}.

Random stellar-age uncertainties were propagated through 500 Monte Carlo realisations using the reported Age16, Age50, and Age84 values. The lower and upper uncertainties were represented by a two-sided Gaussian approximation, after which stars were reassigned to the corresponding age bins and the gradients were recalculated. This test accounts for the reported random age uncertainties but not for possible systematic shifts in the absolute age scale. \rev{The sensitivity of the MGAR to the adopted stellar-age scale is examined separately using an alternative age catalogue in Appendix~\ref{app:age_scale}.}

\subsection{Population and stellar-parameter controls}

To assess whether the full-sample MGAR is affected by the changing mixture of chemical populations, we repeated the gradient measurements separately for stars with $[\alpha/{\rm M}]<0.10$ and $[\alpha/{\rm M}]>0.20$, referred to here as the low-$\alpha$ and high-$\alpha$ populations, respectively. Stars between the two thresholds were excluded from the population-separated measurements. These definitions are used as chemical diagnostics and are not intended to provide a unique decomposition into geometrically defined thin and thick discs \citep{Hayden2017,Feuillet2019,Queiroz2023}.

As a robustness test, we also derived a metallicity-dependent low-/high-$\alpha$ separator from two-component Gaussian-mixture fits to the [$\alpha$/M] distributions in narrow [Fe/H] intervals. Alternative classifications accounting for [$\alpha$/M] uncertainties, together with combined chemical and $Z_{\max}$ selections, are described in Appendix~\ref{Appendix:population_systematics}.

We further tested the effect of the age-dependent $T_{\rm eff}$--$\log g$ distribution in our dwarf- and turn-off-dominated sample. Because the youngest and oldest populations have insufficient common parameter-space coverage, this analysis was restricted to the five central intervals, $2$--$4$, $4$--$6$, $6$--$8$, $8$--$10$, and $10$--$12$ Gyr. The samples were first restricted to their overlapping $T_{\rm eff}$--$\log g$ domain and then compared using two complementary controls.

In the hard-matching analysis, the same number of stars was randomly retained from each age interval within common $T_{\rm eff}$--$\log g$ cells, and the procedure was repeated over 100 realisations. In the reweighted analysis, the cell occupancies were reweighted towards a symmetric target distribution defined by the normalised geometric mean of the occupancies across the five age intervals, with uncertainties estimated from 300 weighted-resampling realisations. All comparison curves were measured using the MWPotential2014 guiding radius over $5.0$--$10.5$ kpc \citep{Bovy2015}. Residual differences in the controlled $T_{\rm eff}$--$\log g$ distributions were quantified using standardised offsets in the mean stellar parameters.

\section{Results}\label{Sec:Results}

\subsection{Age-dependent chemical structure}
\label{SubSec:Age_Chemo_Dynamical}

Before examining the age dependence of the radial metallicity gradient, we first characterise the chemical properties of stellar populations at different ages. This provides the chemical context for assessing whether the changing contributions of low- and high-$\alpha$ populations influence the inferred MGAR.

\begin{figure*}
	\centering
	\includegraphics[width=0.9\textwidth]{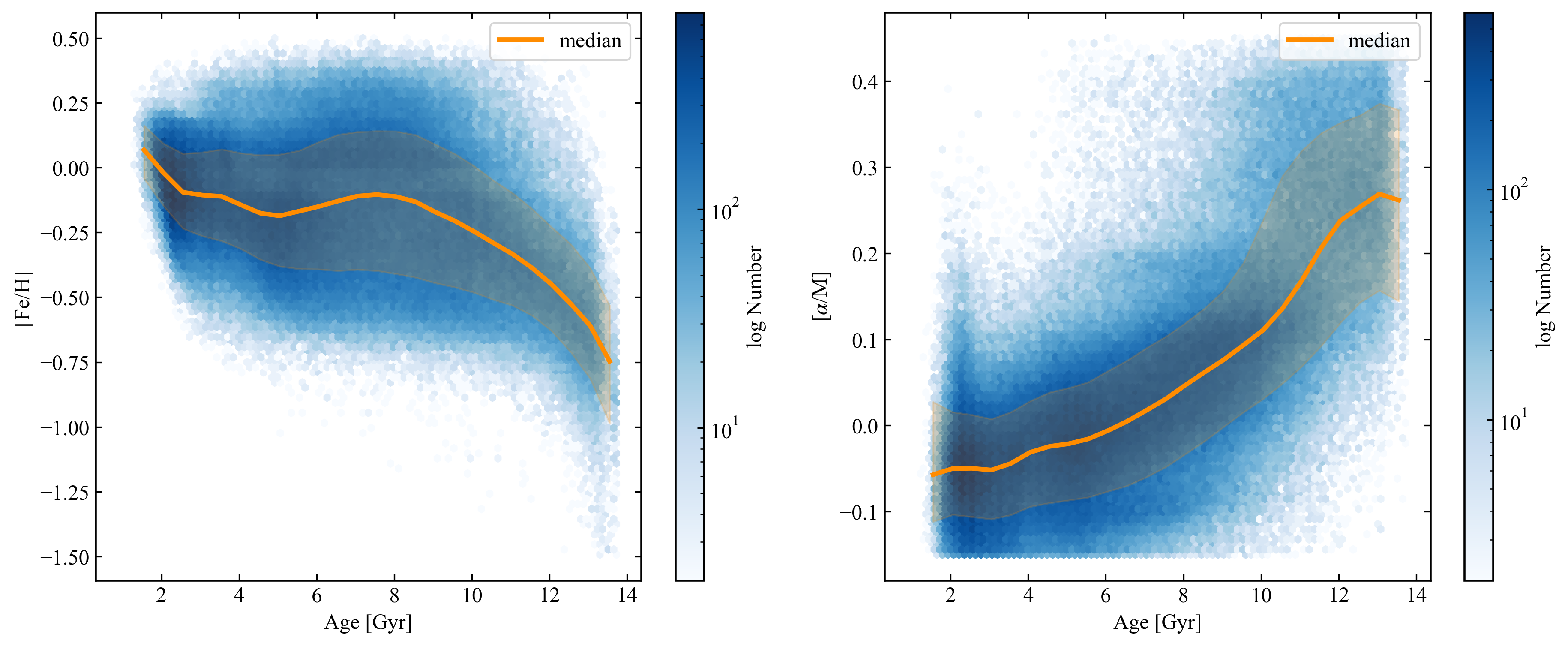}
	\caption{Density distributions of stellar metallicity ([Fe/H], left) and alpha abundance ([$\alpha$/M], right) as a function of stellar age for the solar-neighbourhood sample. The colour scale represents the logarithmic stellar number density. The orange solid curves indicate the median trends, and the shaded regions show the 16th--84th percentile ranges in each age bin.}
	\label{fig:age_chemistry_relations}
\end{figure*}

Fig.~\ref{fig:age_chemistry_relations} shows the distributions of [Fe/H] and [$\alpha$/M] as a function of stellar age for the solar-neighbourhood sample. Both abundance indicators exhibit clear age-dependent trends. The median metallicity decreases towards older populations, from near-solar values at young ages to substantially lower metallicities among the oldest stars, while the metallicity distribution tends to broaden with age. In contrast, the median [$\alpha$/M] abundance increases with stellar age, with a broader distribution among older populations. The sample therefore shows a systematic transition from relatively metal-rich, low-$\alpha$ stars at young ages to more metal-poor, $\alpha$-enhanced stars at old ages. These trends are broadly consistent with previous studies based on large spectroscopic and astrometric samples \citep{Haywood2013,Feuillet2019,Xiang2022}.

\begin{figure*}
	\centering
	\includegraphics[width=0.9\textwidth,height=0.42\textheight,keepaspectratio]{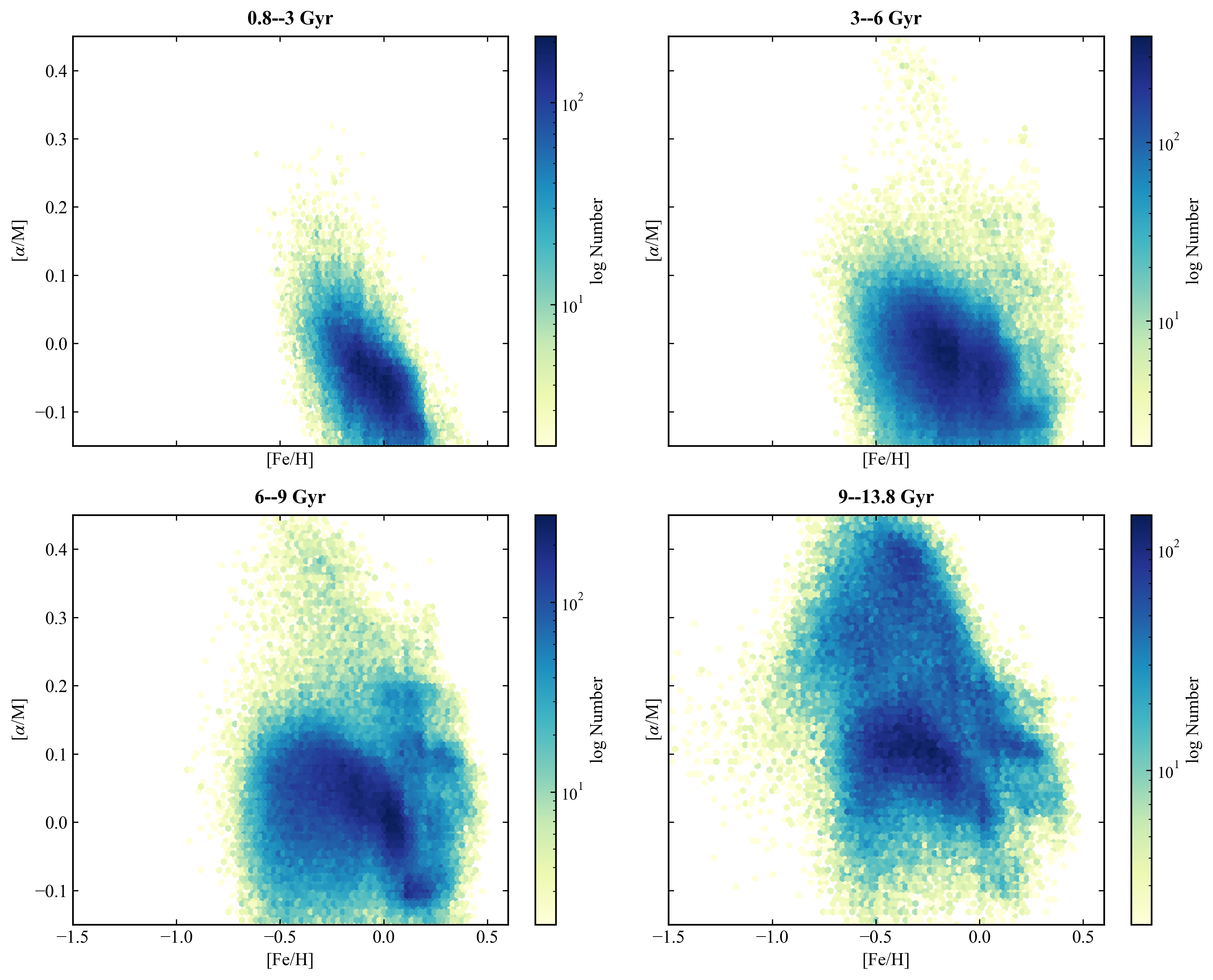}
	\caption{Distribution of stars in the [Fe/H]--[$\alpha$/M] abundance plane for different stellar age bins. The colour scale indicates the logarithmic stellar number density.}
	\label{fig:chemical_plane_in_age_bins}
\end{figure*}

The corresponding evolution of the abundance distribution in the [Fe/H]--[$\alpha$/M] plane is shown in Fig.~\ref{fig:chemical_plane_in_age_bins}. Young stars are concentrated predominantly in the metal-rich, low-$\alpha$ region, whereas progressively older populations extend towards lower metallicities and higher [$\alpha$/M]. The high-$\alpha$ sequence becomes increasingly prominent towards older ages, while young populations are dominated by the low-$\alpha$ sequence. At intermediate ages, however, stars occupy both chemical sequences over a substantial region of the abundance plane. Stellar age therefore does not map uniquely onto either the low- or high-$\alpha$ population, and chemically distinct stars coexist over the same broad age range. Similar age-dependent changes in the abundance distribution have been reported in previous APOGEE and LAMOST studies \citep{Feuillet2019,Xiang2022,Queiroz2023}.

This changing chemical composition is directly relevant to the interpretation of the radial MGAR. Because the relative contributions of low- and high-$\alpha$ populations vary systematically with age, part of the full-sample relation could in principle reflect population mixing rather than the age dependence of the gradient within a single chemical population. In the following section, we first establish the fiducial MGAR for the full sample and then test whether its characteristic intermediate-to-old-age behaviour persists when the low- and high-$\alpha$ populations are considered separately.

\subsection{The radial metallicity gradient--age relation}
\label{SubSec:Radial_metallicity_gradient}

\begin{figure}
	\centering
	\includegraphics[width=0.9\columnwidth]{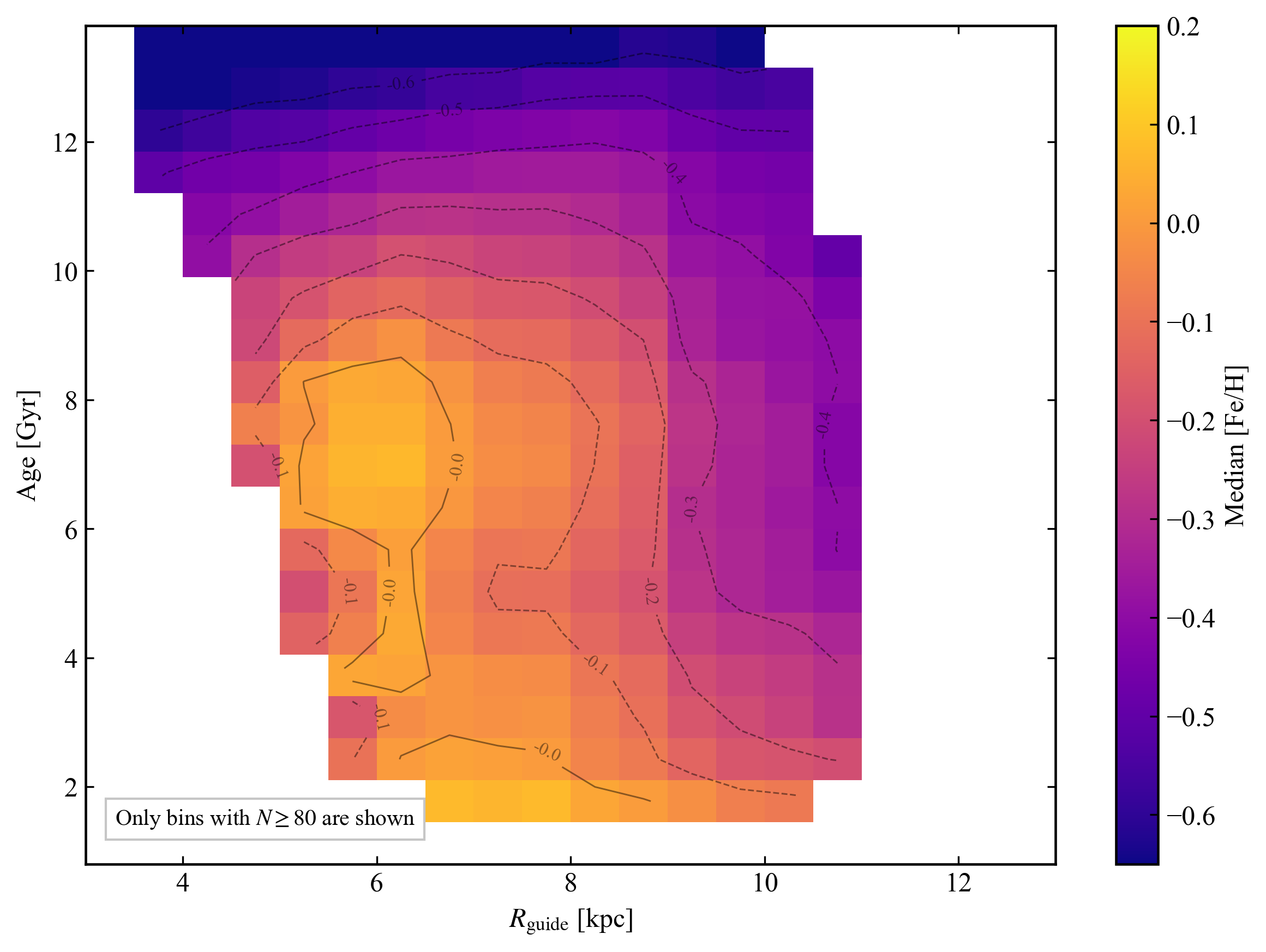}
	\caption{Two-dimensional distribution of the median metallicity in the $R_{\rm guide}$--age plane for the solar-neighbourhood sample. The colour scale indicates the median [Fe/H] value in each bin, while the contour lines trace regions of constant metallicity. Only bins containing at least 80 stars are shown.}
	\label{fig:Rguide_Age_median_feh_heatmap}
\end{figure}

We now quantify how the radial metallicity structure varies with stellar age. Fig.~\ref{fig:Rguide_Age_median_feh_heatmap} presents the two-dimensional distribution of median metallicity in the $R_{\rm guide}$--age plane. At most ages, stars associated with smaller guiding radii are systematically more metal rich than those associated with larger guiding radii, reflecting the well-established negative \rev{stellar radial [Fe/H] gradient} of the Galactic disc \citep{Anders2017,Queiroz2023}. The strength of this radial contrast, however, varies substantially with stellar age. It remains clearly visible among young and intermediate-age populations but becomes markedly weaker towards the oldest ages, indicating that the present-day radial chemical structure traced by the sample is strongly age dependent.

\begin{figure*}
	\centering
	\includegraphics[width=0.9\textwidth,height=0.42\textheight,keepaspectratio]{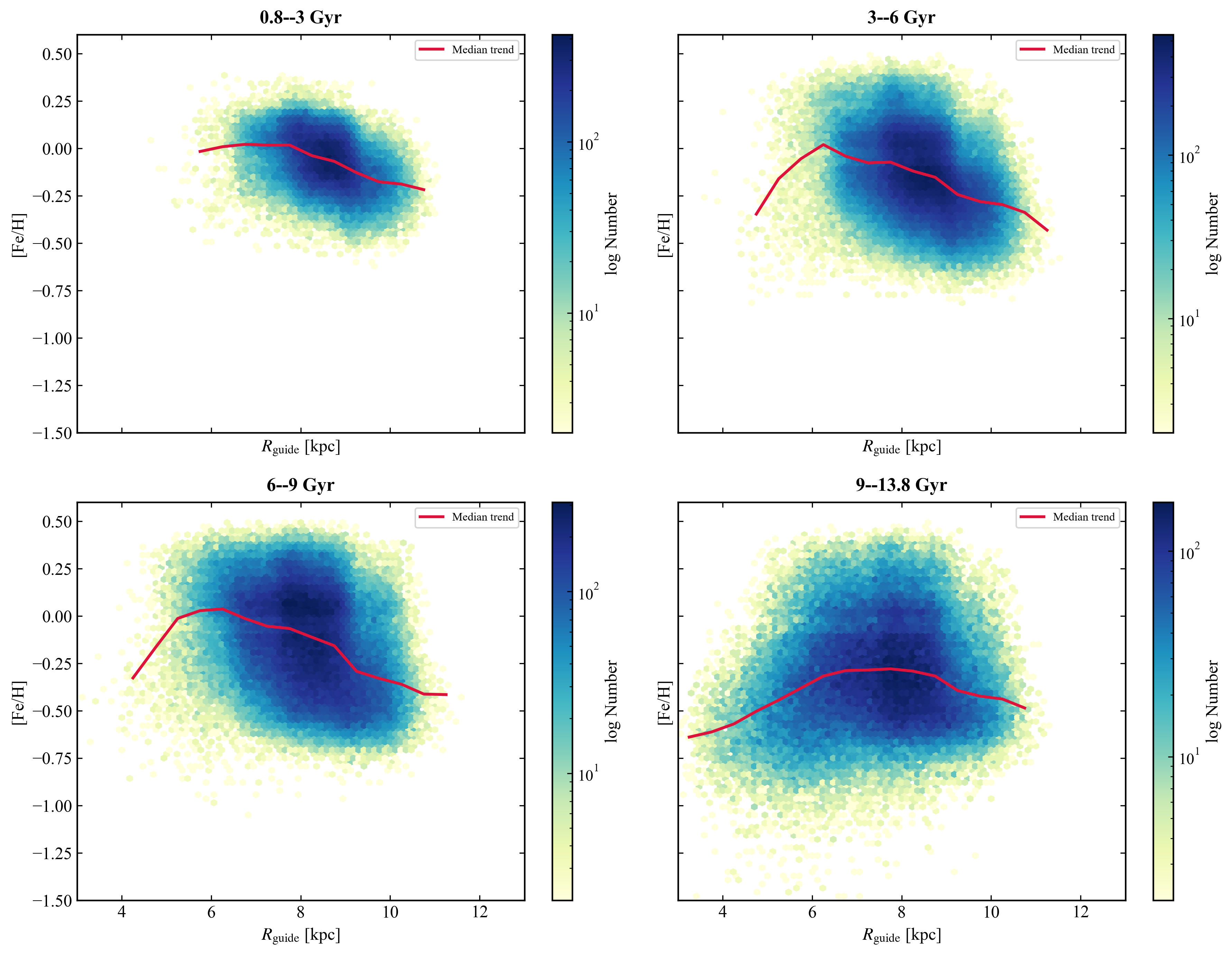}
	\caption{Distribution of stars in the $R_{\rm guide}$--[Fe/H] plane for different stellar age bins. The colour scale represents the logarithmic stellar number density, while the red solid curves indicate the median metallicity trends as a function of guiding radius.}
	\label{fig:feh_Rguide_density_in_age_bins}
\end{figure*}

The same behaviour is shown more directly in Fig.~\ref{fig:feh_Rguide_density_in_age_bins}, which presents the $R_{\rm guide}$--[Fe/H] distributions for representative age intervals. Young populations exhibit relatively narrow metallicity distributions together with clear negative radial trends. The radial metallicity contrast becomes particularly pronounced at intermediate ages, whereas the oldest populations show broader metallicity distributions and substantially weaker radial variations. The oldest age interval also exhibits noticeable curvature in the median radial metallicity profile, indicating that a single global linear slope provides only an approximate description of its radial structure.

\begin{figure}
	\centering
	\includegraphics[width=0.9\columnwidth]{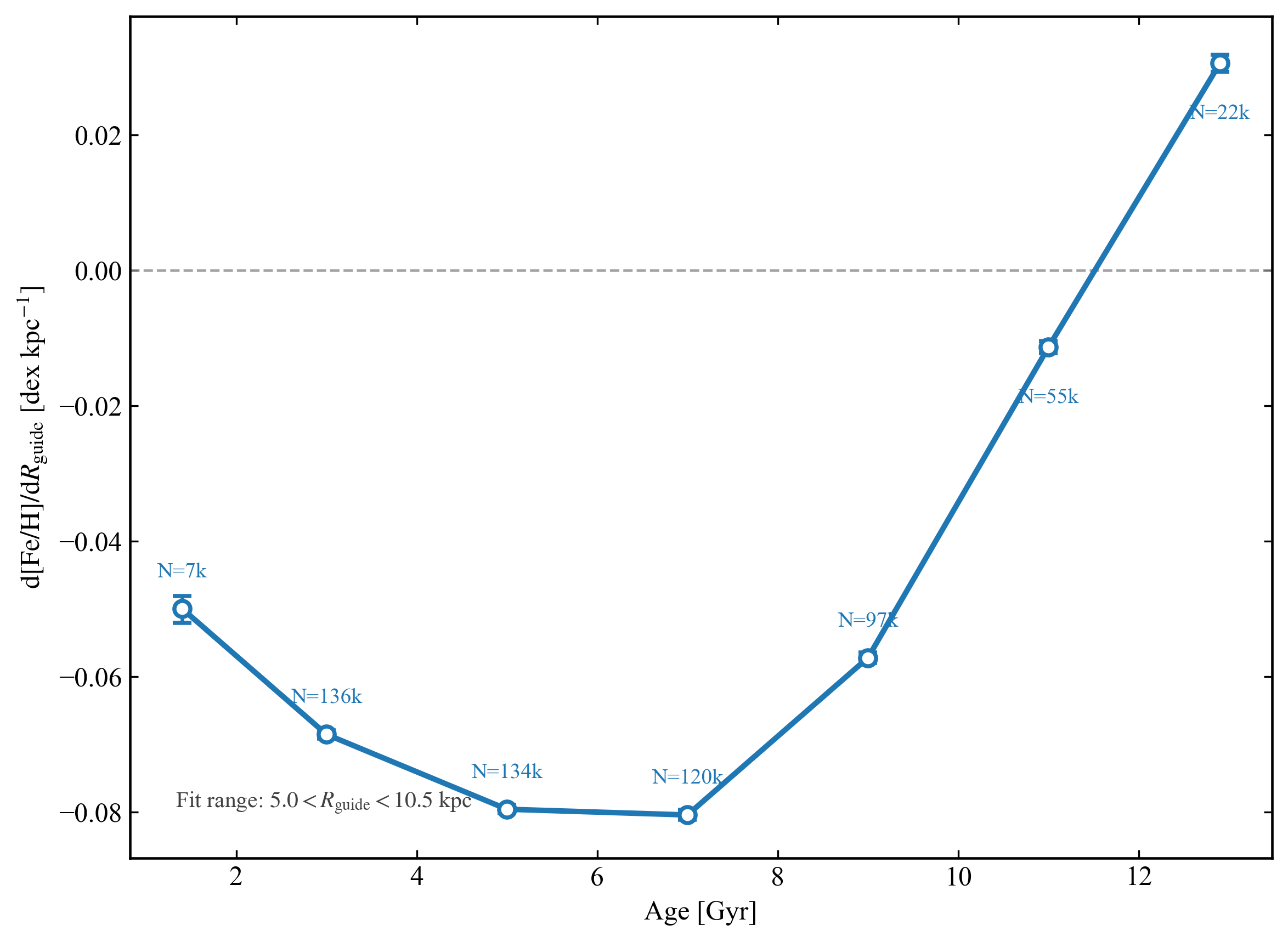}
	\caption{Age dependence of the radial metallicity gradient obtained from the fiducial direct-fit analysis over $5.0<R_{\rm guide}<10.5$ kpc.}
	\label{fig:radial_metallicity_slope_direct_fit}
\end{figure}

We quantify this age dependence by measuring the radial metallicity gradient in independent age bins. The fiducial gradients are obtained from direct linear fits to individual stars over the common interval $5.0<R_{\rm guide}<10.5$ kpc and are shown in Fig.~\ref{fig:radial_metallicity_slope_direct_fit}. The fiducial MGAR is non-monotonic: the gradient becomes more negative from young to intermediate ages and subsequently flattens strongly towards older populations. In the adopted 2-Gyr binning, the most negative gradient occurs in the $6$--$8$ Gyr interval, although the neighbouring $4$--$6$ Gyr bin yields a statistically similar slope. We therefore interpret the fiducial relation as exhibiting a broad intermediate-age turnover rather than a sharply defined minimum at a particular age. The weakly positive global slope obtained for the oldest population is sensitive to the adopted radial interval and age treatment and is therefore not interpreted as evidence for a robust reversal of the Galactic metallicity gradient.

The intermediate-to-old-age flattening is recovered under a range of alternative analysis choices. Fits to binned median metallicity profiles reproduce the same qualitative weakening of the radial gradient towards old ages, although the absolute gradient amplitudes differ in some age bins. Additional tests involving radial fitting ranges, guiding-radius definitions, and vertical-orbit selections are presented in Appendix~\ref{Appendix:gradient_validation}. Although these choices affect the numerical slopes and, in some cases, the location of the most negative age bin, they do not remove the substantial flattening towards older populations. \rev{The flattening is also recovered when the analysis is repeated using an alternative stellar-age scale, although the quantitative gradients at intermediate ages are more sensitive to the adopted age estimator (Appendix~\ref{app:age_scale}).}

We next examine whether the fiducial MGAR could arise primarily from the changing mixture of chemically distinct populations. As an initial diagnostic separation, stars with $[\alpha/{\rm M}]<0.10$ are assigned to a low-$\alpha$ population and those with $[\alpha/{\rm M}]>0.20$ to a high-$\alpha$ population, while stars in the intermediate region are excluded. These selections are intended to isolate chemically distinct populations and should not be interpreted as a unique decomposition into geometrically
defined thin and thick discs. Chemically defined low- and high-$\alpha$ populations can overlap substantially in kinematic space \citep{Jing2026}, and chemical and geometric disc definitions need not map uniquely onto one
another on a star-by-star basis
\citep{Hayden2017,Feuillet2019,Queiroz2023}.

\begin{figure}
	\centering
	\includegraphics[width=0.9\columnwidth]{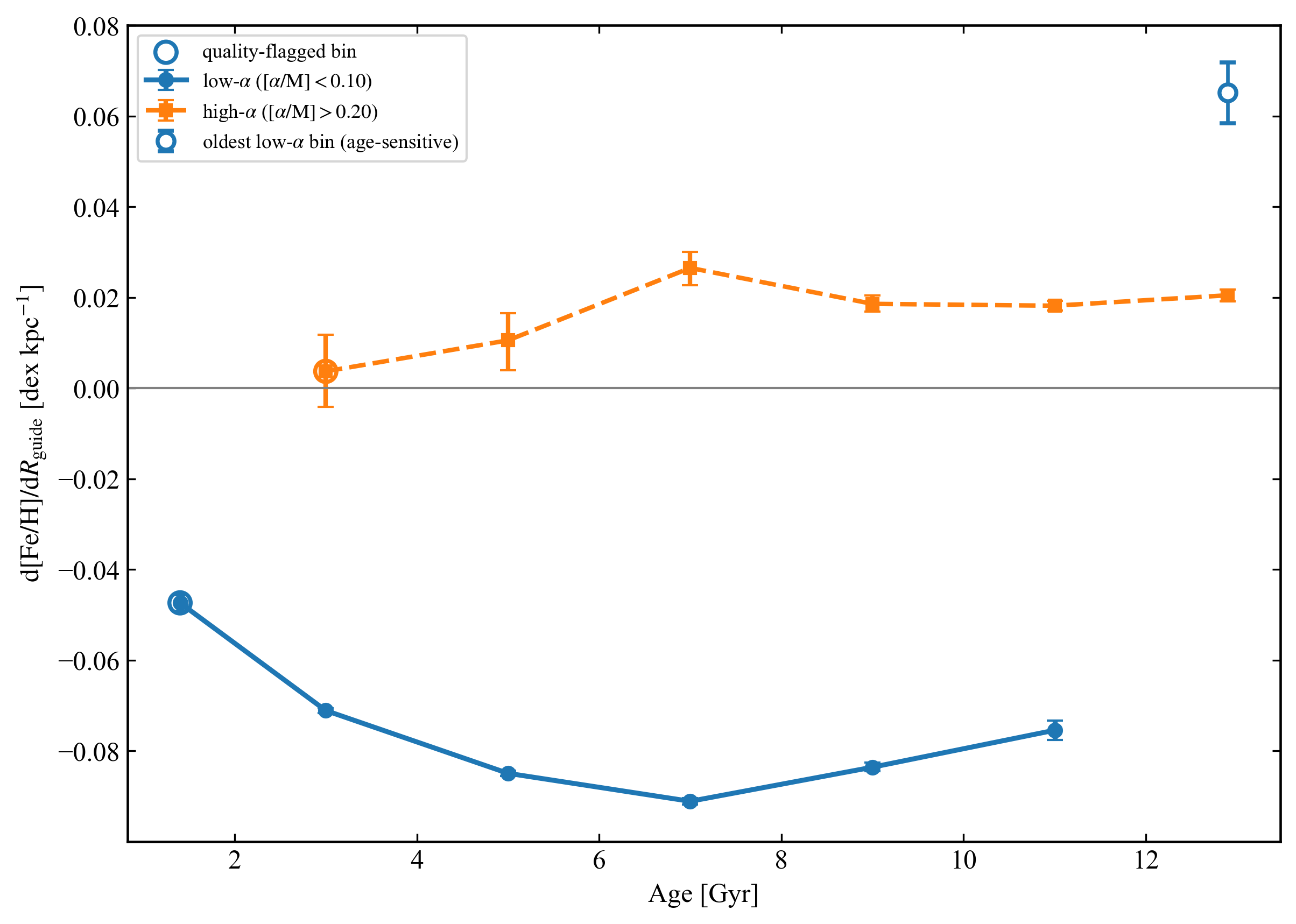}
	\caption{Age dependence of the radial metallicity gradient for the fiducial chemically defined low- and high-$\alpha$ populations, selected using $[\alpha/{\rm M}]<0.10$ and $[\alpha/{\rm M}]>0.20$, respectively. Error bars indicate the bootstrap uncertainties. Open symbols identify bins that do not satisfy the adopted sample-size or radial-coverage quality criteria. The oldest low-$\alpha$ bin is shown separately because its inferred gradient is particularly sensitive to stellar-age uncertainties and the upper age boundary.}
	\label{fig:alpha_population_gradient_fiducial}
\end{figure}

As shown in Fig.~\ref{fig:alpha_population_gradient_fiducial}, the low-$\alpha$ population retains a clear age dependence, with relatively strong negative gradients at intermediate ages followed by substantial flattening towards older populations. In contrast, the high-$\alpha$ population exhibits considerably flatter gradients over the age range in which it is sufficiently represented, broadly consistent with previous measurements of chemically $\alpha$-enhanced disc populations \citep{Hayden2014,Imig2023}. The persistence of the intermediate-to-old-age flattening within the low-$\alpha$ population indicates that the full-sample MGAR cannot be explained solely by the increasing contribution of high-$\alpha$ stars towards older ages. The oldest low-$\alpha$ bin is particularly sensitive to stellar-age uncertainties and the adopted upper age boundary, and its positive slope is therefore not interpreted as evidence for a robust gradient reversal. Alternative fixed abundance boundaries and metallicity-dependent Gaussian-mixture classifications lead to the same qualitative conclusion and are discussed in Appendix~\ref{Appendix:population_systematics}.

Chemical-population mixing is therefore insufficient by itself to account for the observed MGAR morphology. A separate concern arises from the age dependence of the stellar-parameter distribution in a dwarf- and turn-off-dominated sample: different age bins occupy systematically different regions of the $T_{\rm eff}$--$\log g$ plane, so part of the measured MGAR variation could reflect changes in the tracer population rather than age alone. Because the youngest and oldest populations have insufficient overlapping support in this plane, the controlled comparison is restricted to five central 2-Gyr age bins spanning $2$--$12$ Gyr, with bin centres at 3, 5, 7, 9, and 11 Gyr. Within these bins, we first restrict the samples to their overlapping two-dimensional stellar-parameter domain and then apply two stronger controls: symmetric reweighting and hard matching of the $T_{\rm eff}$--$\log g$ distributions.

\begin{figure*}
	\centering
	\includegraphics[width=0.9\textwidth]{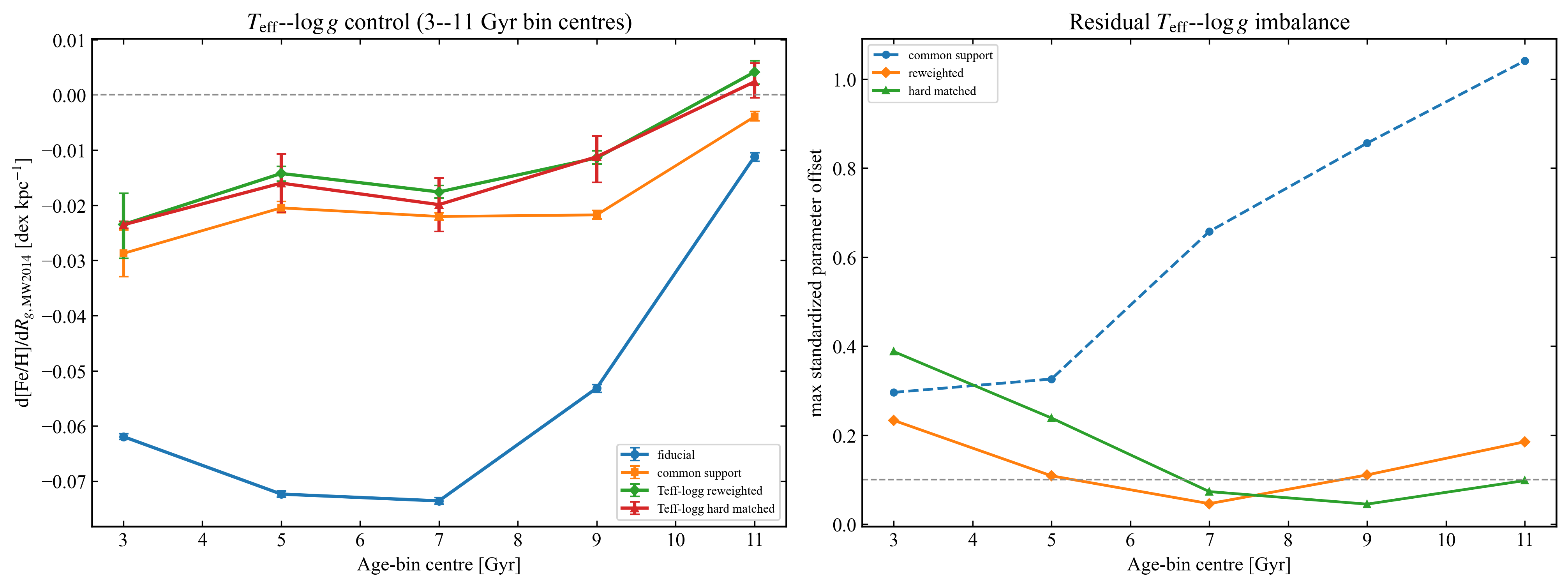}
	\caption{Effect of controlling the age-dependent $T_{\rm eff}$--$\log g$ distribution on the MGAR for five central 2-Gyr age bins spanning $2$--$12$ Gyr, with bin centres at 3, 5, 7, 9, and 11 Gyr. Left: gradients measured for the uncontrolled sample and after restricting to the overlapping $T_{\rm eff}$--$\log g$ domain, symmetric reweighting, and hard matching of the stellar-parameter distributions. All gradients in this comparison are measured using $R_{g,\mathrm{MWPotential2014}}$. Right: maximum absolute standardised offset of the mean $T_{\rm eff}$ or $\log g$ relative to the pooled adjusted target for the controlled samples. The horizontal dashed line at 0.1 is shown only as a reference level for residual imbalance rather than as a formal threshold.}
	\label{fig:teff_logg_control}
\end{figure*}

The resulting MGARs are shown in Fig.~\ref{fig:teff_logg_control}. Restricting the analysis to the overlapping stellar-parameter domain already reduces the absolute gradient amplitudes substantially relative to the corresponding uncontrolled relation. The reweighted and hard-matched samples yield similarly shallower gradients, demonstrating that the measured MGAR amplitude is sensitive to the age-dependent stellar-parameter distribution. The behaviour towards the younger end of the controlled interval is also modified: in particular, the progressive steepening towards the intermediate-age minimum seen in the uncontrolled relation is no longer recovered in the same form once the stellar-parameter distributions are controlled. The detailed behaviour at younger ages should therefore not be regarded as independent of stellar-parameter selection.

Despite these changes, the controlled relations continue to show an overall flattening from the intermediate-age regime towards older populations within the $2$--$12$ Gyr comparison range. A weak local structure remains around the intermediate-age bins, but its detailed form is much less stable than the subsequent flattening. The right-hand panel of Fig.~\ref{fig:teff_logg_control} quantifies the remaining imbalance in $T_{\rm eff}$ and $\log g$. Restricting the analysis to the overlapping domain alone does not equalise the stellar-parameter distributions, whereas both reweighting and hard matching substantially improve the balance, particularly over the intermediate-age range. Residual differences remain, especially towards the younger end of the controlled interval, and the matched and reweighted relations should therefore not be interpreted as completely free of stellar-parameter selection effects.

\rev{Because spectroscopic metallicity systematics may themselves depend on stellar atmospheric parameters, we also tested the controlled MGAR using APOGEE DR17 metallicities for stars in common with LAMOST. Keeping the stellar sample, ages, guiding radii, and $T_{\rm eff}$--$\log g$ control fixed, the APOGEE-based MGAR closely reproduces the broad age dependence obtained with LAMOST metallicities. This comparison indicates that the main features of the controlled MGAR are relatively insensitive to the adopted spectroscopic [Fe/H] scale, although it does not directly establish whether the changes induced by the stellar-parameter control are reproduced using APOGEE metallicities. The detailed comparison is presented in Appendix~\ref{Appendix:population_systematics}.}

Taken together, these analyses show that the gradient amplitudes and the detailed behaviour at younger ages are appreciably affected by the stellar-parameter distribution, whereas the flattening from intermediate ages through the $10$--$12$ Gyr population is substantially more stable under changes in chemical-population definition, radial coordinate, vertical-orbit selection, propagation of the reported random age uncertainties, \rev{the adopted stellar-age scale,} radial fitting range, and $T_{\rm eff}$--$\log g$ distribution. The precise location and depth of the intermediate-age minimum are therefore less robust, while the behaviour of the oldest $12$--$13.8$ Gyr bin is particularly sensitive to the adopted radial interval and other analysis choices and is not used to infer a robust gradient reversal. We therefore identify the intermediate-to-old-age flattening as the most stable empirical feature of the observed MGAR.

\subsection{Dynamical context of the age-dependent radial structure}\label{SubSec:stellar_kinematics}

To place this age-dependent radial chemical structure in a dynamical context, we examine the kinematic properties of the age-selected stellar populations. The age--velocity dispersion relation provides a standard characterisation of the dynamical temperature of Galactic disc populations \citep{Aumer2009,Aumer2016,YuLiu2018,Mackereth2019,Sharma2021}.

\begin{figure}
	\centering
	\includegraphics[width=0.9\columnwidth]{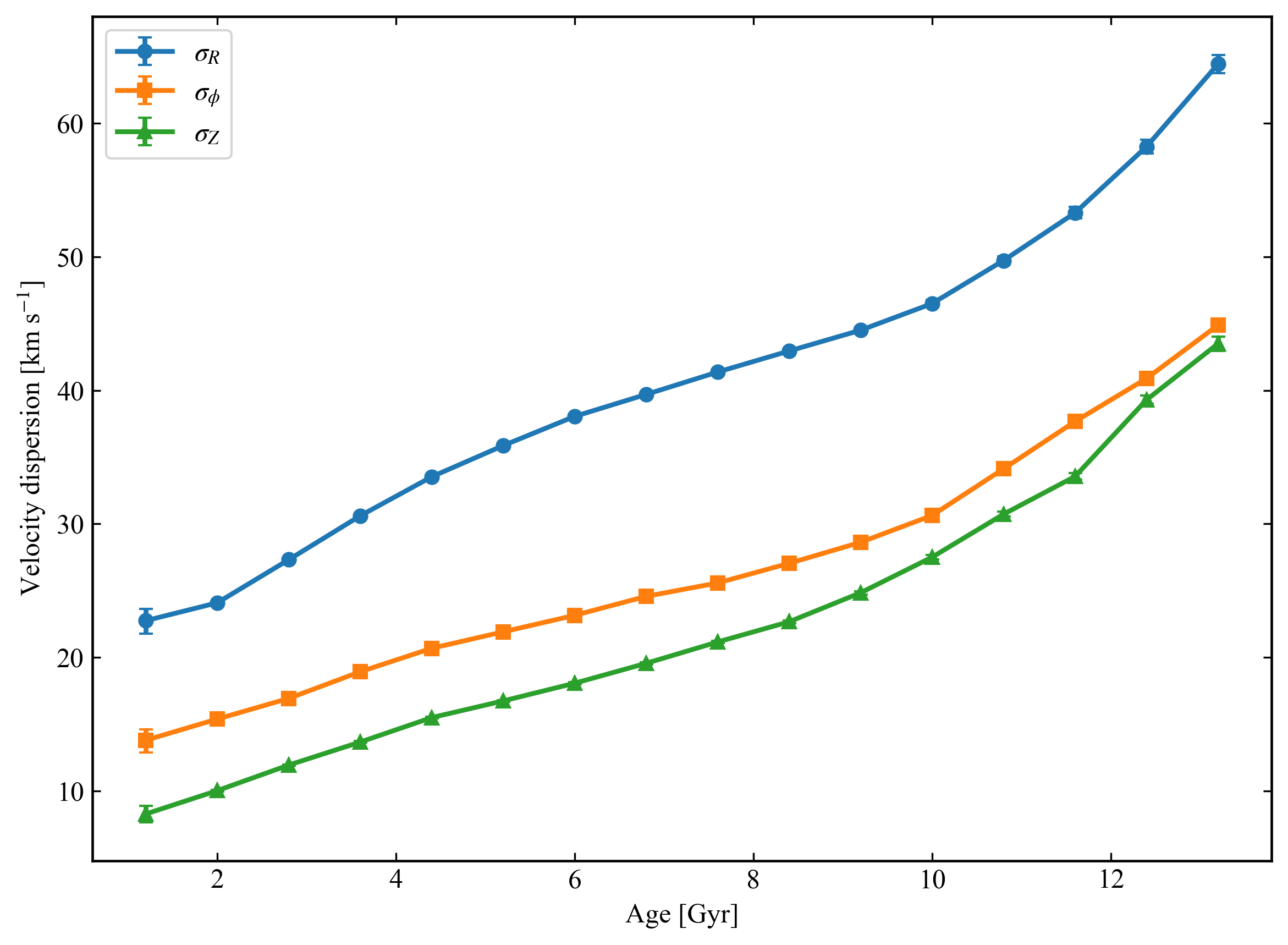}
	\caption{Radial ($\sigma_R$), azimuthal ($\sigma_\phi$), and vertical ($\sigma_Z$) velocity dispersions as a function of stellar age for the solar-neighbourhood sample. Error bars indicate the uncertainties of the velocity dispersions in each age bin.}\label{fig:velocity_dispersion_vs_age}
\end{figure}

Fig.~\ref{fig:velocity_dispersion_vs_age} shows the radial ($\sigma_R$), azimuthal ($\sigma_\phi$), and vertical ($\sigma_Z$) velocity dispersions as a function of stellar age. All three components increase systematically towards older populations, indicating that older stars are dynamically hotter. The radial dispersion remains the largest over most of the age range, while the azimuthal and vertical components show the same overall increase with age.

\begin{figure}
	\centering
	\includegraphics[width=0.9\columnwidth]{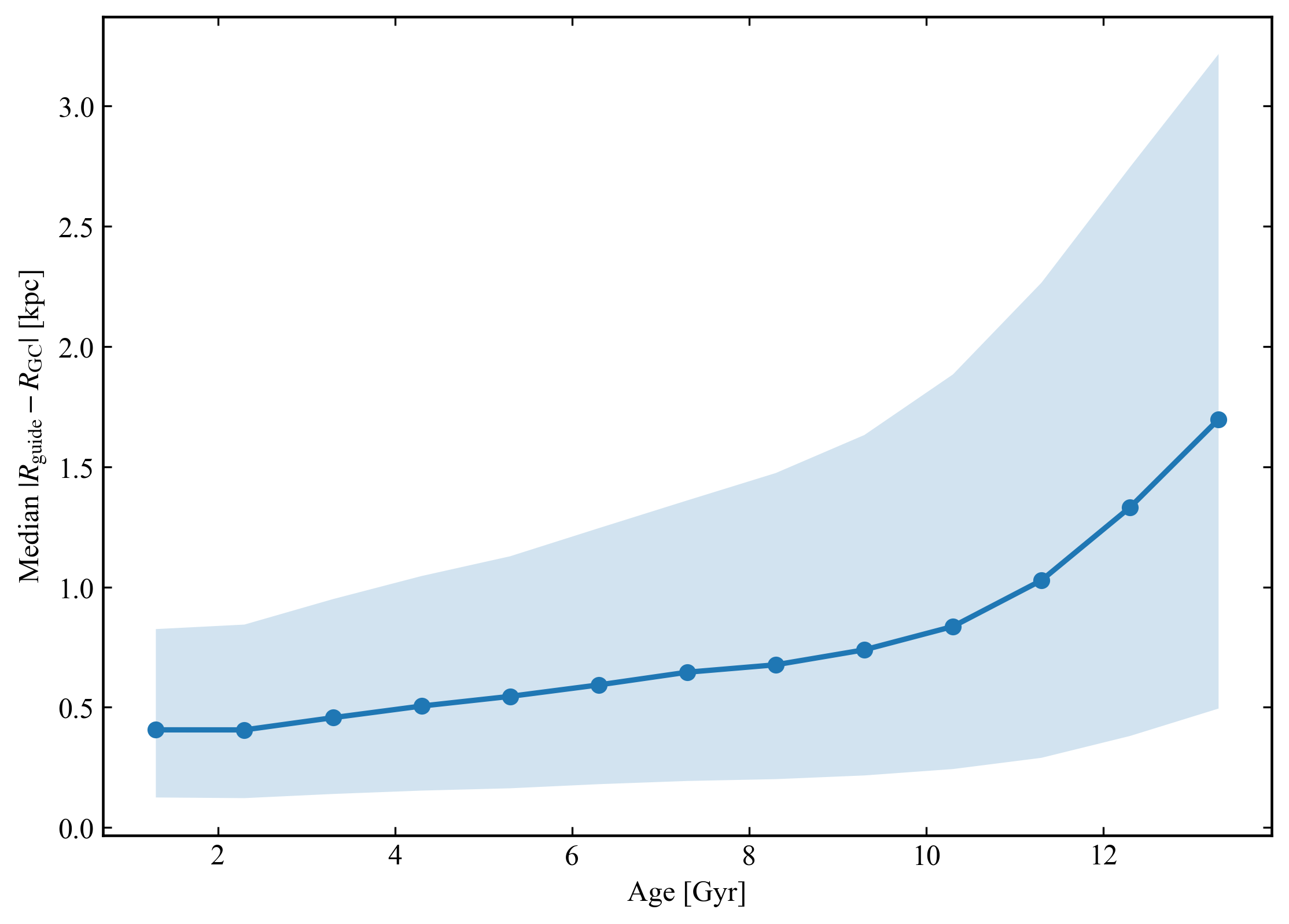}
	\caption{Median absolute difference between the guiding radius and the instantaneous Galactocentric radius, $|R_{\rm guide}-R_{\rm GC}|$, as a function of stellar age. The shaded region indicates the 16th--84th percentile range. The increase towards older populations reflects progressively larger present-day radial orbital excursions.}\label{fig:radial_orbital_excursion_age}
\end{figure}

Fig.~\ref{fig:radial_orbital_excursion_age} shows the median absolute difference between the guiding radius and the instantaneous Galactocentric radius, $|R_{\rm guide}-R_{\rm GC}|$, as a function of stellar age. This quantity increases systematically with age, showing that older stars are, on average, observed farther from their guiding radii. The trend is consistent with the larger epicyclic amplitudes expected for dynamically hotter populations.

We emphasize that $|R_{\rm guide}-R_{\rm GC}|$ is an instantaneous measure of a star's displacement from its guiding centre and therefore primarily traces radial orbital excursions, or blurring, rather than changes in guiding radius or angular momentum associated with churning \citep{Sellwood2002,Schonrich2009,Minchev2018}. The increase in dynamical temperature and radial orbital excursions towards older ages occurs alongside the weakening of the radial metallicity gradient, but their common dependence on stellar age does not establish a causal connection. The present analysis therefore cannot determine the relative contributions of birth-gradient evolution, orbital blurring, churning, and other forms of secular dynamical evolution to the observed MGAR. These possibilities and the corresponding limitations are discussed in Section~\ref{Sec:Discussion}.

\section{Discussion}\label{Sec:Discussion}

\subsection{The diversity of observed metallicity gradient--age relations}

Published measurements of the MGAR do not converge on a single morphology. Main-sequence turn-off and subgiant samples have shown relatively shallow gradients among the youngest and oldest stars, with stronger negative gradients at intermediate ages \citep{Xiang2015,Wang2019}. Non-monotonic behaviour has also been found in red-giant samples \citep{Anders2017,Anders2023}, whereas other analyses based on guiding radii, asteroseismic giant samples, or red-clump populations favour a smoother flattening towards older ages \citep{Vickers2021,Willett2023,Chen2026,Sun2026}. The published diversity therefore does not follow a simple division by tracer class and may instead reflect a combination of differences in stellar tracers, age determination, spatial coverage, abundance calibration, radial-coordinate definition, chemical selection, and survey selection function.

Our fiducial dwarf- and turn-off-dominated sample likewise exhibits a non-monotonic MGAR, but its detailed morphology is sensitive to the stellar-parameter distribution sampled by the adopted tracer selection. Controlling the age-dependent $T_{\rm eff}$--$\log g$ distribution substantially reduces the absolute gradient amplitudes and modifies the younger-age branch of the relation, whereas the flattening from intermediate to older ages remains considerably more stable. The detailed non-monotonic morphology of the fiducial MGAR should therefore not be regarded as equally robust at all ages; in particular, the intermediate-to-old-age flattening is more stable than the absolute gradient amplitudes and the detailed behaviour at younger ages.

Stellar-tracer and parameter-selection effects can therefore contribute to the diversity of published MGARs, consistent with previous evidence that inferred radial metallicity gradients vary across stellar populations occupying different regions of parameter space \citep{Yu2012,Wang2023a}. They cannot, however, provide a unique explanation for the literature differences, because both non-monotonic and smoother relations have been recovered from giant-star samples. Since the present analysis is not a controlled dwarf--giant comparison under matched spatial coverage, chemical selection, and age calibration, differences in survey selection, spatial sampling, abundance scales, and stellar ages may remain equally important.

\subsection{Robustness of the observed gradient--age relation}

In the fiducial analysis, the most negative gradient occurs in the $6$--$8$ Gyr interval, but the neighbouring $4$--$6$ Gyr bin has a statistically similar slope. Propagation of the reported random stellar-age uncertainties and changes in the analysis choices can shift the most negative gradient between neighbouring intermediate-age bins. We therefore interpret the result as a broad intermediate-age turnover rather than a sharply defined physical transition at a specific age.

\rev{The comparison with an alternative stellar-age scale provides an additional test of the sensitivity of the MGAR to the adopted ages. Using the LAMOST--Gaia subgiant ages of \citet{Xiang2022}, with the updated catalogue presented by \citet{Xiang2025}, the broad age dependence of the MGAR is recovered, including the substantial flattening from intermediate to older ages (Appendix~\ref{app:age_scale}). The quantitative gradients differ most strongly at intermediate ages, particularly around $5$--$9$ Gyr, where substantial redistribution between age bins is also present. We therefore regard the intermediate-to-old-age flattening as more stable with respect to the adopted age scale than the detailed gradient amplitudes and intermediate-age morphology.}

The chemical-population tests further show that the observed age dependence is not generated solely by the changing mixture of low- and high-$\alpha$ populations. The low-$\alpha$ population retains relatively strong negative gradients at intermediate ages followed by substantial flattening towards older populations, whereas the high-$\alpha$ population exhibits considerably flatter gradients where it is sufficiently represented, broadly consistent with previous measurements \citep{Hayden2014,Vickers2021,Imig2023}. This conclusion remains unchanged under alternative chemical classifications and when the low-$\alpha$ population is further restricted to vertically confined orbits. Chemical and vertical population mixing are therefore insufficient by themselves to explain the intermediate-to-old-age flattening.

The oldest populations require particular caution because their median [Fe/H]--radius relation is not well represented by a single linear gradient, and the inferred slope depends appreciably on the adopted radial interval and population selection. Mildly positive slopes recovered in some configurations should therefore not be interpreted as evidence for a robust reversal of the Galactic metallicity gradient; a more conservative interpretation is that the radial metallicity structure becomes weak and radially non-linear at the oldest ages. Taken together, the robustness tests indicate that the strongest empirical result is the substantial flattening from intermediate to older populations, whereas the precise location and depth of the intermediate-age minimum, the absolute gradient amplitudes, and the detailed younger-age morphology are less stable.

\subsection{Chemo-dynamical implications and limitations}

The present-day stellar [Fe/H] MGAR is not a direct record of the temporal evolution of the gas-phase abundance gradient. \rev{In addition to the dynamical effects discussed below, a direct chemical identification is not appropriate. Stellar metallicity studies commonly use Fe as a tracer, whereas gas-phase metallicity in Galactic H\,{\sc ii} regions is generally characterised using O/H \citep{MendezDelgado2022}. Because O is produced predominantly on short timescales by massive stars, whereas Fe receives an important delayed contribution from Type Ia supernovae, the two elements trace different enrichment histories, and their radial abundance gradients need not evolve identically \citep{MendezDelgado2024}. Comparisons between H\,{\sc ii} regions and planetary nebulae provide complementary constraints on the temporal evolution of nebular abundance gradients, but these O/H-based relations should likewise not be directly identified with the stellar [Fe/H] MGAR \citep{Amayo2026}.} The metallicity distribution of a present-day mono-age stellar population reflects both the radial chemical structure of the interstellar medium at formation and the subsequent evolution of stellar orbits \citep{Sellwood2002,Schonrich2009,Minchev2013}. Chemical-evolution models allow the \rev{gas-phase abundance gradient at the time of stellar birth} itself to evolve with time \citep{Matteucci1989,Chiappini1997,Matteucci2001,Sharda2021a}, while subsequent stellar orbital evolution can further modify the relation between metallicity and present-day radius. Present-day Galactocentric or guiding radii therefore cannot generally be identified with stellar birth radii, as also illustrated by attempts to reconstruct birth radii and birth metallicity gradients \citep{Minchev2018,Ratcliffe2023,Lu2024}. Cosmological simulations likewise indicate that part of the present-day age dependence may already be imprinted at birth rather than being generated entirely by subsequent orbital evolution \citep{Graf2025a}.

The kinematic trends do not remove this physical degeneracy. Although dynamically hotter populations with larger present-day radial orbital excursions coexist with weaker metallicity gradients at older ages, their common dependence on stellar age does not establish a causal connection. In particular, $|R_{\rm guide}-R_{\rm GC}|$ provides a proxy for the present-day epicyclic displacement associated with orbital blurring, rather than for changes in guiding radius or angular momentum associated with churning \citep{Sellwood2002,Schonrich2009,Minchev2013}. 
The present analysis therefore cannot uniquely separate \rev{the chemical-abundance structure imprinted at stellar birth} from subsequent orbital evolution through churning and blurring, or from broader secular dynamical evolution. Distinguishing these contributions will require independent birth-radius constraints or forward chemo-dynamical modelling.

Several observational limitations should also be kept in mind. The present sample is not volume complete, and the LAMOST targeting strategy together with the spectroscopic, age-quality, and colour--magnitude selections can introduce correlations between stellar age, metallicity, and atmospheric parameters. The $T_{\rm eff}$--$\log g$ controls address one component of this selection but do not constitute a full correction for the survey selection function. \rev{The same-star comparison with APOGEE DR17 further shows that the broad age dependence of the controlled MGAR is recovered on an external metallicity scale. Although mild parameter-dependent offsets remain between the LAMOST and APOGEE [Fe/H] measurements, the main features of the controlled MGAR are relatively insensitive to the adopted abundance scale. Nevertheless, this comparison does not fully constrain the contribution of spectroscopic abundance systematics to the changes induced by the stellar-parameter control.} \rev{The Monte Carlo experiments propagate the reported random StarHorse age uncertainties, while the comparison with an alternative age catalogue provides an empirical test of the sensitivity to the adopted age scale. Nevertheless, neither test constitutes a direct propagation of the full age--metallicity covariance, which cannot be reconstructed from the marginal posterior summaries available in the public StarHorse catalogue.} Finally, the measured gradients describe metallicity as a function of present-day guiding radius for stars observed in the solar neighbourhood and should not be interpreted as a direct spatial census of the Galactic disc over the fitted radial interval.

The principal empirical constraint from this study is therefore that the detailed MGAR morphology is partly sensitive to the stellar-parameter distribution sampled by the adopted tracer selection, whereas the flattening from intermediate to older populations is substantially more stable. Determining the physical origin of this flattening will require matched tracer samples and selection functions, independent age calibrations and birth-radius constraints, and forward chemo-dynamical modelling.

\section{Conclusions}\label{Sec:Conclusion}

Using a solar-neighbourhood sample of FGK dwarf and turn-off stars constructed from LAMOST DR7 spectroscopy, Gaia DR3 astrometry, and stellar ages from the corrected StarHorse catalogue, we investigated the MGAR of the Galactic disc and its chemical and dynamical context. Our main conclusions are as follows.

\begin{enumerate}
	
	\item The fiducial MGAR is non-monotonic, with relatively strong negative gradients at intermediate ages followed by pronounced flattening towards older populations. In the adopted 2-Gyr age bins, the most negative gradient occurs at $6$--$8$ Gyr, but the neighbouring $4$--$6$ Gyr bin has a statistically similar slope. We therefore interpret the relation as showing a broad intermediate-age turnover rather than a sharply defined transition at a particular age. The mildly positive gradient recovered for the oldest stars is sensitive to the radial fitting range and population selection and is not regarded as evidence for a robust reversal of the Galactic metallicity gradient.
	
    \item The intermediate-to-old-age flattening is the most robust feature of the observed MGAR. It is recovered under alternative guiding-radius definitions, radial fitting ranges, vertical-orbit selections, propagation of the reported random stellar-age uncertainties, \rev{an alternative stellar-age scale,} and chemical-population definitions. In contrast, controlling the age-dependent $T_{\rm eff}$--$\log g$ distribution by restricting the samples to an overlapping stellar-parameter domain and applying reweighting and hard matching substantially reduces the absolute gradient amplitudes and modifies the detailed behaviour at younger ages. \rev{A same-star comparison using APOGEE DR17 [Fe/H] measurements recovers the broad age dependence of the controlled MGAR, indicating that its main features are relatively insensitive to the adopted spectroscopic abundance scale.} The gradient amplitude and young-age morphology are therefore more sensitive to the stellar-parameter distributions of the populations entering the different age bins, whereas the subsequent flattening towards older ages is substantially more stable.
	
	\item The age dependence of the full-sample MGAR cannot be explained solely by the changing mixture of low- and high-$\alpha$ populations. The low-$\alpha$ population retains relatively strong negative gradients at intermediate ages followed by substantial flattening towards older populations, whereas the high-$\alpha$ population exhibits considerably flatter gradients over the age range in which it is sufficiently represented. The same qualitative conclusion is obtained using metallicity-dependent Gaussian-mixture classifications and when the low-$\alpha$ population is additionally restricted to vertically confined orbits. Chemical and vertical population mixing therefore \rev{cannot by themselves account for} the intermediate-to-old-age flattening.
	
	\item The chemical and kinematic properties of the sampled populations also vary systematically with stellar age. Older stars are generally more metal poor and $\alpha$ enhanced, dynamically hotter, and characterised by larger present-day radial orbital excursions as indicated by $|R_{\rm guide}-R_{\rm GC}|$. This quantity provides a proxy for the present-day epicyclic displacement associated with orbital blurring, rather than for changes in guiding radius or angular momentum associated with churning. These trends provide a chemo-dynamical context for the observed MGAR but do not establish a causal origin for the intermediate-to-old-age flattening. The present data cannot uniquely separate the \rev{imprint of the radial chemical-abundance structure at stellar birth} from subsequent orbital evolution through churning and blurring, or from the broader secular dynamical evolution of the disc.
	
\end{enumerate}

Overall, distinguishing the selection-sensitive features of the MGAR from the more stable intermediate-to-old-age flattening may help reconcile apparently discrepant gradient--age relations reported in previous studies and provides an empirical constraint for chemo-dynamical models of Galactic disc evolution. Future comparisons of dwarf and giant tracers under matched spatial coverage and selection functions, together with consistent abundance scales and independent age calibrations, will be important for determining how much of the observed MGAR diversity reflects genuine Galactic evolution and how much arises from differences in the sampled stellar populations.

\section*{Acknowledgements}

This work was supported by the Kunlun Talents Program of Qinghai Province (Grant No.~2025-QLGKLYCZX-026) and by the Qinghai University Research Ability Enhancement Project (Grant No.~2025KTSQ26). This work also used data products from the Guoshoujing Telescope (the Large Sky Area Multi-Object Fiber Spectroscopic Telescope, LAMOST; \url{https://www.lamost.org}). LAMOST is a National Major Scientific Project built by the Chinese Academy of Sciences. Funding for the project has been provided by the National Development and Reform Commission. LAMOST is operated and managed by the National Astronomical Observatories, Chinese Academy of Sciences.

This research has also made use of data from the European Space Agency (ESA) mission Gaia (\url{https://www.cosmos.esa.int/gaia}), processed by the Gaia Data Processing and Analysis Consortium (DPAC, \url{https://www.cosmos.esa.int/web/gaia/dpac/consortium}). Funding for the DPAC has been provided by national institutions, in particular the institutions participating in the Gaia Multilateral Agreement. We further made use of stellar ages and distances provided by the corrected StarHorse catalogue of \citet{Queiroz2023}, which combines Gaia DR3 astrometry and photometry with spectroscopic survey information within a Bayesian framework. \rev{For the alternative age-scale test, we used the stellar-age catalogue of \citet{Xiang2022}, adopting the updated catalogue presented by \citet{Xiang2025}. We also made use of SDSS-IV APOGEE-2 DR17 data. Funding for the Sloan Digital Sky Survey IV has been provided by the Alfred P. Sloan Foundation, the U.S. Department of Energy Office of Science, and the Participating Institutions. SDSS-IV acknowledges support and resources from the Center for High-Performance Computing at the University of Utah. The SDSS web site is \url{https://www.sdss4.org}. SDSS is managed by the Astrophysical Research Consortium for the Participating Institutions of the SDSS Collaboration, which are listed at \url{https://www.sdss4.org/collaboration/affiliations/}.}

\section*{Data Availability}

The data underlying this article are publicly available. The LAMOST DR7 data can be accessed from the LAMOST data release website at \url{https://www.lamost.org}. Gaia data are available from the ESA Gaia archive at \url{https://gea.esac.esa.int/archive/}. The stellar ages and distances used in the fiducial analysis were obtained from the corrected StarHorse catalogue presented by \citet{Queiroz2023}, which is publicly available through the CDS/VizieR service. \rev{The alternative stellar ages used for the age-scale robustness test were obtained from the catalogue of \citet{Xiang2022}, adopting the updated catalogue presented by \citet{Xiang2025}. The APOGEE DR17 data used for the external metallicity-scale comparison are publicly available as part of SDSS Data Release 17 \citep{Abdurrouf2022}.}



\bibliographystyle{mnras}
\bibliography{references} 




\newpage

\appendix

\section{Robustness of the metallicity gradient--age relation}\label{Appendix:gradient_validation}

The analysis in Section~\ref{SubSec:Radial_metallicity_gradient} shows that the radial metallicity gradient varies systematically with stellar age. In the fiducial analysis, the gradient reaches its most negative values over an intermediate-age regime and subsequently becomes substantially flatter towards older populations. The numerical gradient values and the detailed morphology of the MGAR, however, may depend on the adopted radial coordinate, radial fitting interval, vertical-orbit selection, and stellar-age uncertainties. We therefore examine these effects here to identify which features of the observed MGAR remain stable when the principal analysis choices are varied.

\begin{figure*}
	\centering
	\includegraphics[width=0.9\textwidth]{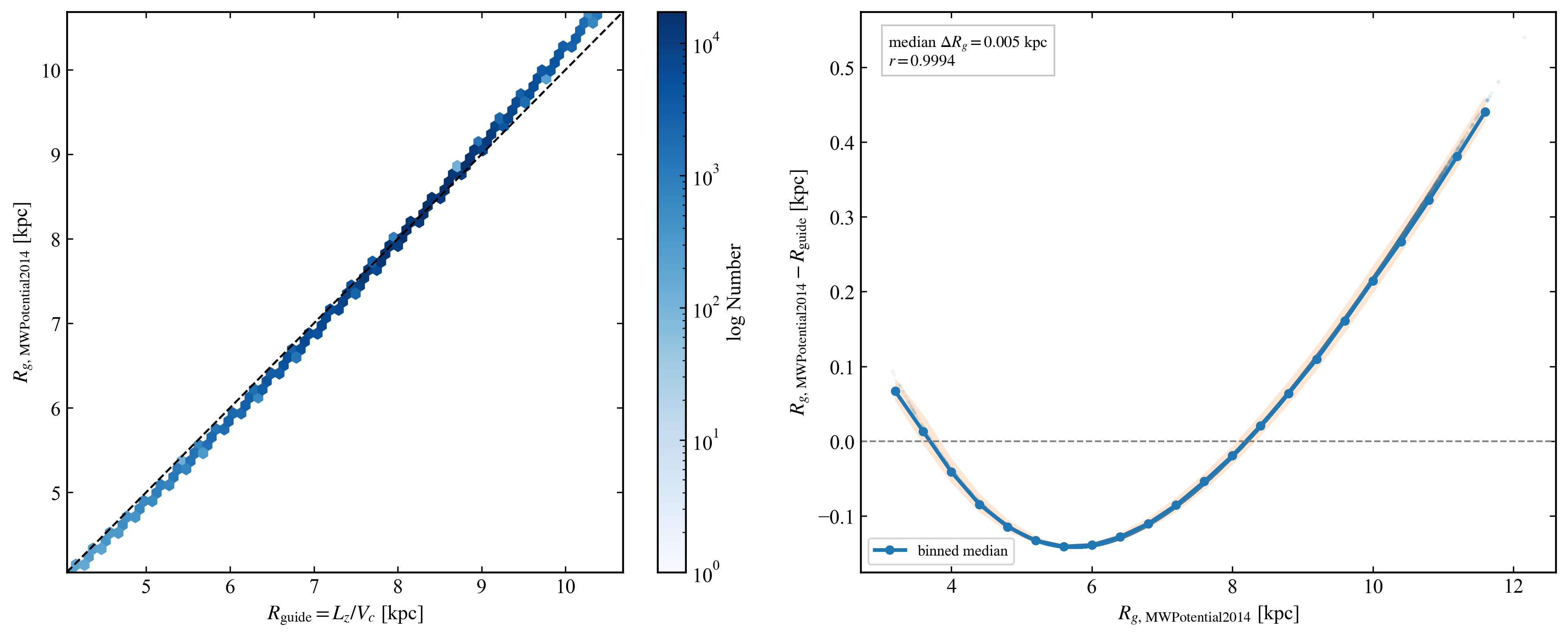}
	\caption{Comparison between the fiducial guiding radius, $R_{\rm guide}=L_Z/V_{\rm circ}$, and the guiding radius $R_{g,\mathrm{MW}}$ calculated in MWPotential2014 \citep{Bovy2015}. Left: direct comparison of the two guiding-radius estimates for the same stars, with the one-to-one relation shown for reference. Right: difference $R_{g,\mathrm{MW}}-R_{\rm guide}$ as a function of $R_{g,\mathrm{MW}}$, together with the binned median trend.}
	\label{fig:appendix_Rguide_comparison}
\end{figure*}

We first assess whether the result depends on the adopted definition of guiding radius. The fiducial analysis uses the approximation $R_{\rm guide}=L_Z/V_{\rm circ}$, whereas an alternative guiding radius, $R_{g,\mathrm{MW}}$, is calculated in MWPotential2014 \citep{Bovy2015}. Fig.~\ref{fig:appendix_Rguide_comparison} compares the two estimates for the same stars. They follow an almost one-to-one relation over the radial range relevant to the present analysis, with only small systematic differences as a function of radius. Such deviations are expected because the fiducial approximation adopts a constant circular velocity, whereas the circular-velocity curve in MWPotential2014 varies with Galactocentric radius.

\begin{figure*}
	\centering
	\includegraphics[width=0.9\textwidth]{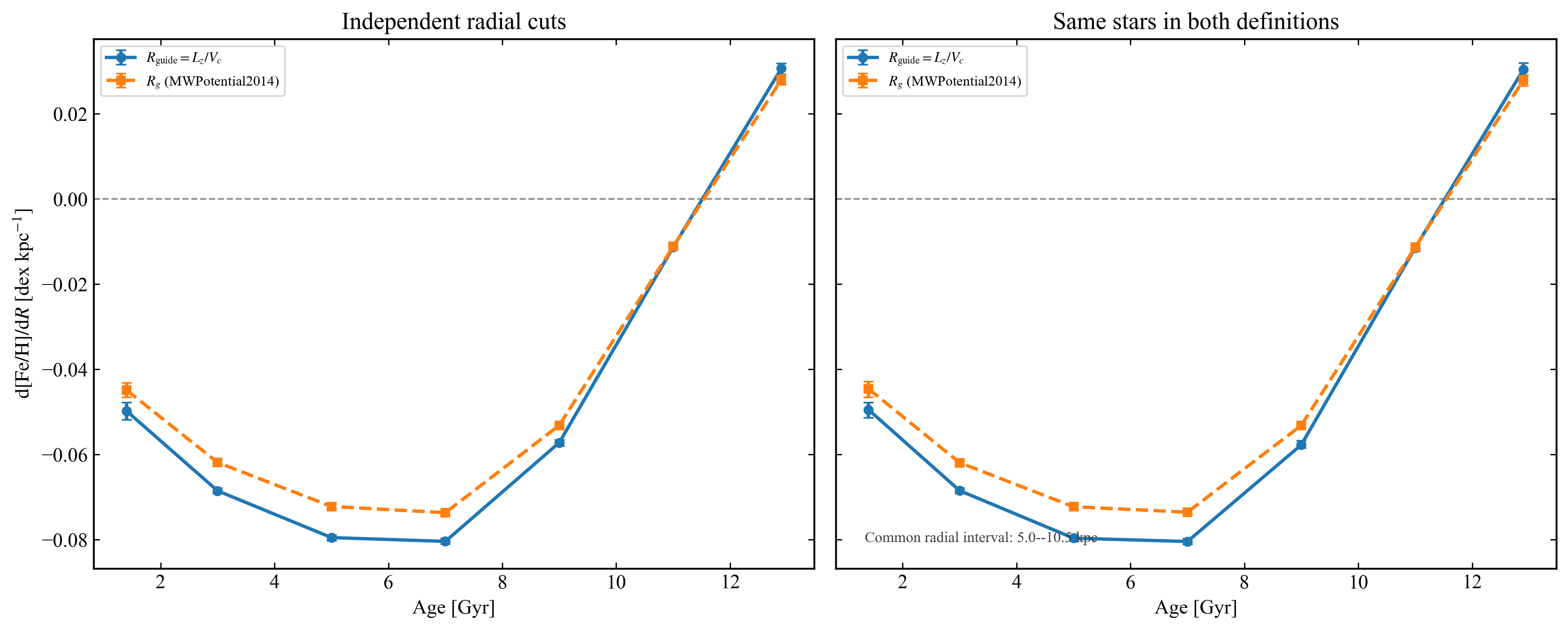}
	\caption{Comparison of the MGARs obtained using the fiducial guiding radius $R_{\rm guide}=L_Z/V_{\rm circ}$ and the guiding radius $R_{g,\mathrm{MW}}$ calculated in MWPotential2014 \citep{Bovy2015}. Left: each radial coordinate is independently restricted to the adopted fitting interval before the gradients are measured. Right: both gradients are measured using the same stars satisfying the adopted radial interval under both guiding-radius definitions.}
	\label{fig:appendix_MGAR_Rguide_MW}
\end{figure*}

The corresponding MGARs are shown in Fig.~\ref{fig:appendix_MGAR_Rguide_MW}. In the left-hand panel, the adopted radial interval is applied independently to each guiding-radius definition, so the two fits can contain slightly different stars near the radial boundaries. In the right-hand panel, the comparison is repeated using only stars that satisfy the adopted radial interval under both definitions, thereby isolating the effect of the radial-coordinate definition from differences in sample membership. The two guiding-radius definitions produce small offsets in the numerical slopes but closely similar MGAR morphologies, and both retain the substantial flattening from intermediate to older populations. The behaviour of the oldest population is less stable and is not interpreted as evidence for a robust reversal of the radial metallicity gradient. The same-star comparison therefore shows that the intermediate-to-old-age flattening is not produced by the simplified $L_Z/V_{\rm circ}$ definition of guiding radius.

\begin{figure*}
	\centering
	\includegraphics[width=0.9\textwidth]{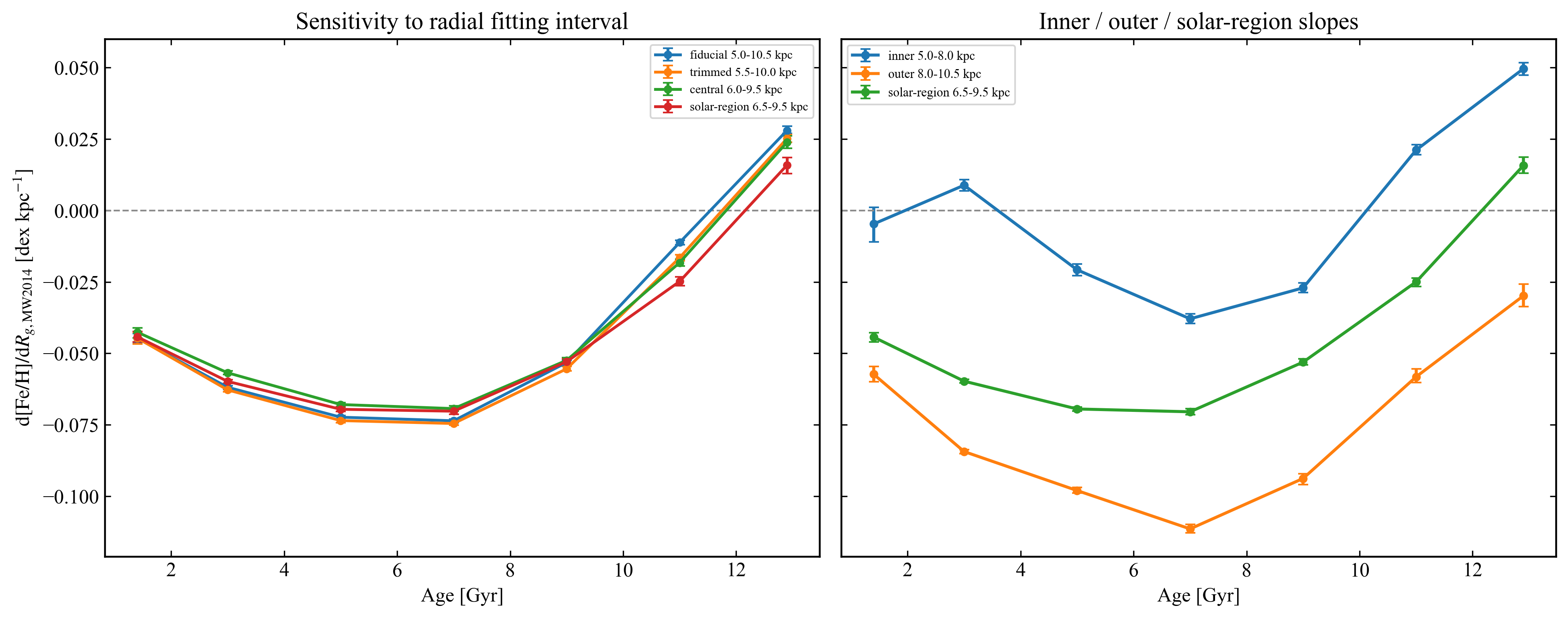}
	\caption{Sensitivity of the MGAR to the adopted radial interval. The figure compares gradients measured over different guiding-radius fitting ranges and over representative radial regions.}
	\label{fig:appendix_radial_range}
\end{figure*}

We next examine whether the inferred MGAR is sensitive to the radial interval used for the linear fit. Fig.~\ref{fig:appendix_radial_range} compares the gradient--age relations obtained over several radial baselines within the region adequately sampled by the data. The absolute slopes vary with the fitting interval, but the same broad pattern is recovered: relatively strong negative gradients occur over the intermediate-age regime and are followed by substantial flattening towards older ages. The persistence of this behaviour over progressively restricted radial ranges shows that the principal intermediate-to-old-age flattening is not determined by a single choice of radial baseline.

The same figure also compares gradients measured over different radial regions. The differences become particularly important for the oldest populations, for which the local slopes measured in the inner, solar-radius, and outer regions are not mutually consistent with a single linear radial profile. This behaviour is consistent with the curvature visible in the oldest $R_{\rm guide}$--[Fe/H] distribution in Fig.~\ref{fig:feh_Rguide_density_in_age_bins}. Consequently, the mildly positive global gradient obtained in some configurations depends on how different portions of this curved profile are weighted. We therefore interpret the oldest population as exhibiting a weak and radially non-linear metallicity structure rather than a robustly positive Galactic metallicity gradient.

A further concern is that the vertical orbital structure of the sample changes systematically with stellar age, so older age bins may contain a larger contribution from vertically extended populations with different chemical properties \citep{Hayden2017,Vickers2021,Chen2026}. To examine this effect, we repeat the gradient analysis after separating stars according to their maximum orbital height, $Z_{\max}$, calculated in MWPotential2014. Unlike the instantaneous vertical position $|Z|$, $Z_{\max}$ characterises the vertical extent of an orbit and is therefore less dependent on the present orbital phase.

\begin{figure*}
	\centering
	\includegraphics[width=0.9\textwidth]{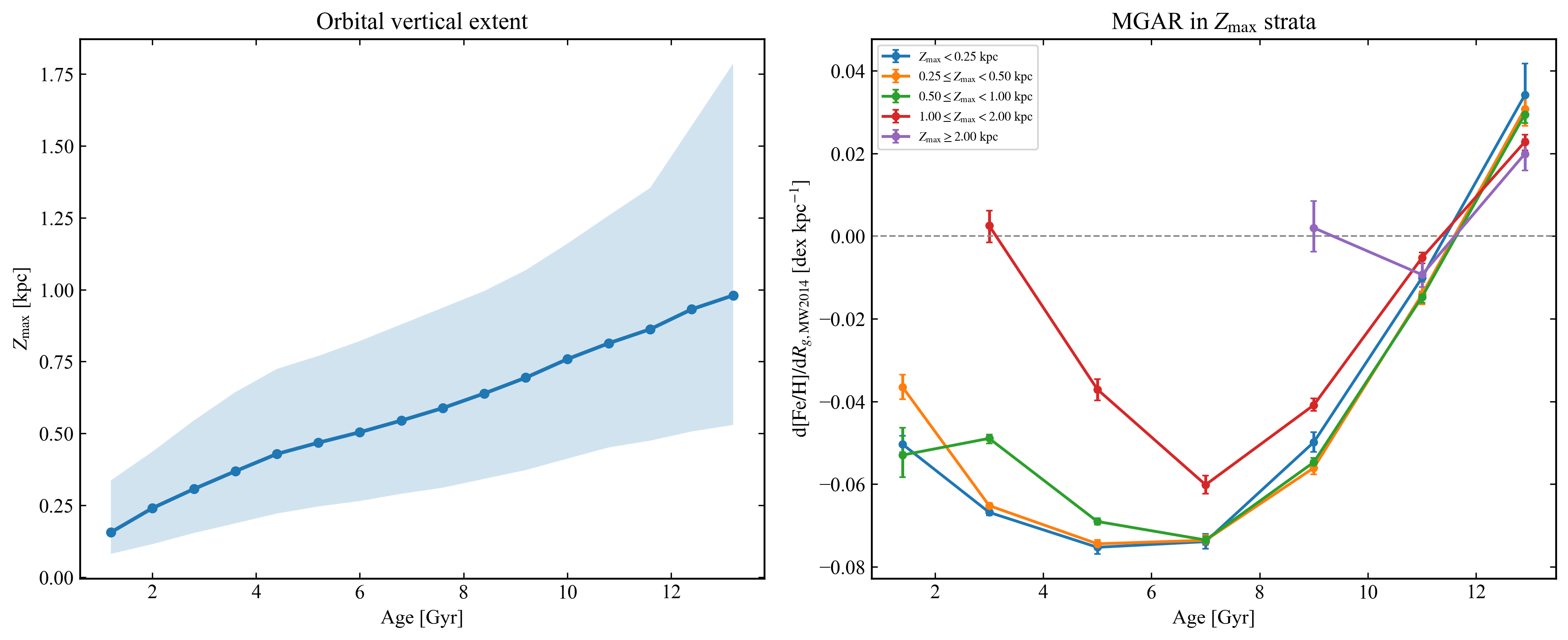}
	\caption{Dependence of the MGAR on vertical orbital structure. The figure shows $Z_{\max}$ as a function of stellar age and the corresponding MGARs measured in different $Z_{\max}$ intervals. The intermediate-to-old-age flattening remains visible even for the vertically confined population with $Z_{\max}<0.25$ kpc.}
	\label{fig:appendix_Zmax}
\end{figure*}

Fig.~\ref{fig:appendix_Zmax} shows the age dependence of $Z_{\max}$ together with the MGAR measured in different $Z_{\max}$ intervals. The gradient amplitudes vary with vertical orbital structure, but the substantial flattening from intermediate to older populations remains visible after the sample is divided by $Z_{\max}$. In particular, the same broad behaviour is retained among the most vertically confined stars with $Z_{\max}<0.25$ kpc. The observed intermediate-to-old-age flattening therefore cannot be explained solely by an increasing contribution of vertically extended populations towards older ages.

Finally, we assess the effect of the reported random stellar-age uncertainties. The individual asymmetric uncertainties are propagated through Monte Carlo realisations using the reported 16th, 50th, and 84th percentiles of the stellar-age posterior. In each realisation, the stellar ages are perturbed and the stars are reassigned to the corresponding age bins, after which the radial metallicity gradients are recalculated using the same radial range and fitting procedure as in the fiducial analysis. This test propagates the reported random age uncertainties but does not address systematic offsets in the absolute stellar-age scale.

\begin{figure}
	\centering
	\includegraphics[width=0.9\columnwidth]{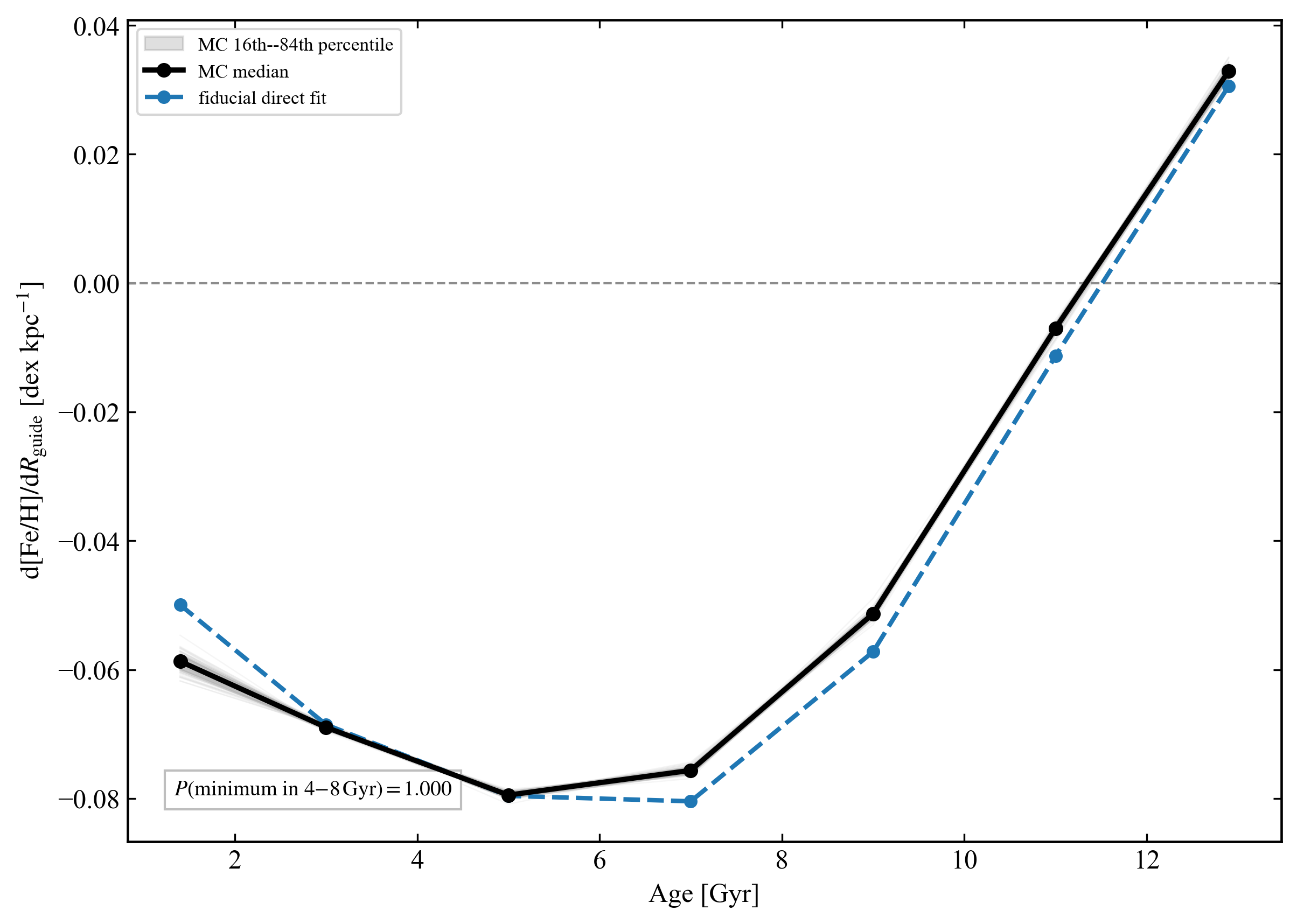}
	\caption{Monte Carlo assessment of the effect of the reported random stellar-age uncertainties on the MGAR. The black curve shows the median gradient obtained from the Monte Carlo realisations, the grey shaded region indicates the 16th--84th percentile interval, and the blue dashed curve shows the fiducial direct-fit relation.}
	\label{fig:appendix_age_MC}
\end{figure}

Fig.~\ref{fig:appendix_age_MC} shows the resulting gradient--age relation. The age perturbations introduce additional scatter in the recovered gradients and allow the most negative individual age bin to shift between neighbouring intermediate-age intervals. In particular, the $4$--$6$ and $6$--$8$ Gyr populations are not cleanly distinguishable as sharply separate extrema, supporting the interpretation of the fiducial minimum as a broad intermediate-age feature rather than a transition localised at a single age. More importantly, the substantial flattening from intermediate to older populations remains present after propagation of the reported random stellar-age uncertainties.

These tests show that the numerical gradient values and the detailed morphology of the MGAR are not invariant to all analysis choices. The location of the most negative gradient can shift between neighbouring intermediate-age bins, while the global slope inferred for the oldest population is particularly sensitive to the adopted radial fitting range because its radial metallicity profile is not well described by a single linear relation. Nevertheless, the substantial flattening from intermediate to older populations is consistently recovered under alternative guiding-radius definitions, radial fitting intervals, vertical-orbit selections, and propagation of the reported random stellar-age uncertainties. We therefore regard this intermediate-to-old-age flattening as more robust than either the precise location and depth of the fiducial minimum or the behaviour of the oldest individual age bin.

\section{\rev{Sensitivity to the adopted stellar-age scale}}\label{app:age_scale}

\rev{As described in Section~\ref{Sec:Data_Age}, metallicity information enters the StarHorse age inference for the LAMOST solutions, and the stellar ages and LAMOST [Fe/H] values used to construct the MGAR therefore cannot be assumed to be statistically independent. The public StarHorse catalogue used here provides marginal posterior summaries for age and metallicity rather than samples from their joint posterior distribution, so the full age--metallicity covariance cannot be reconstructed directly. The Monte Carlo experiment presented in Appendix~\ref{Appendix:gradient_validation} provides a complementary assessment of the impact of the reported marginal age uncertainties by resampling the stellar ages, reassigning stars to age bins, and repeating the gradient measurements. The principal age dependence of the MGAR, particularly the flattening from intermediate to old ages, remains present under this resampling. This experiment propagates the reported marginal age uncertainties but does not constitute a direct propagation of the full age--metallicity covariance.}

\begin{figure*}
	\centering
	\includegraphics[width=0.9\textwidth]{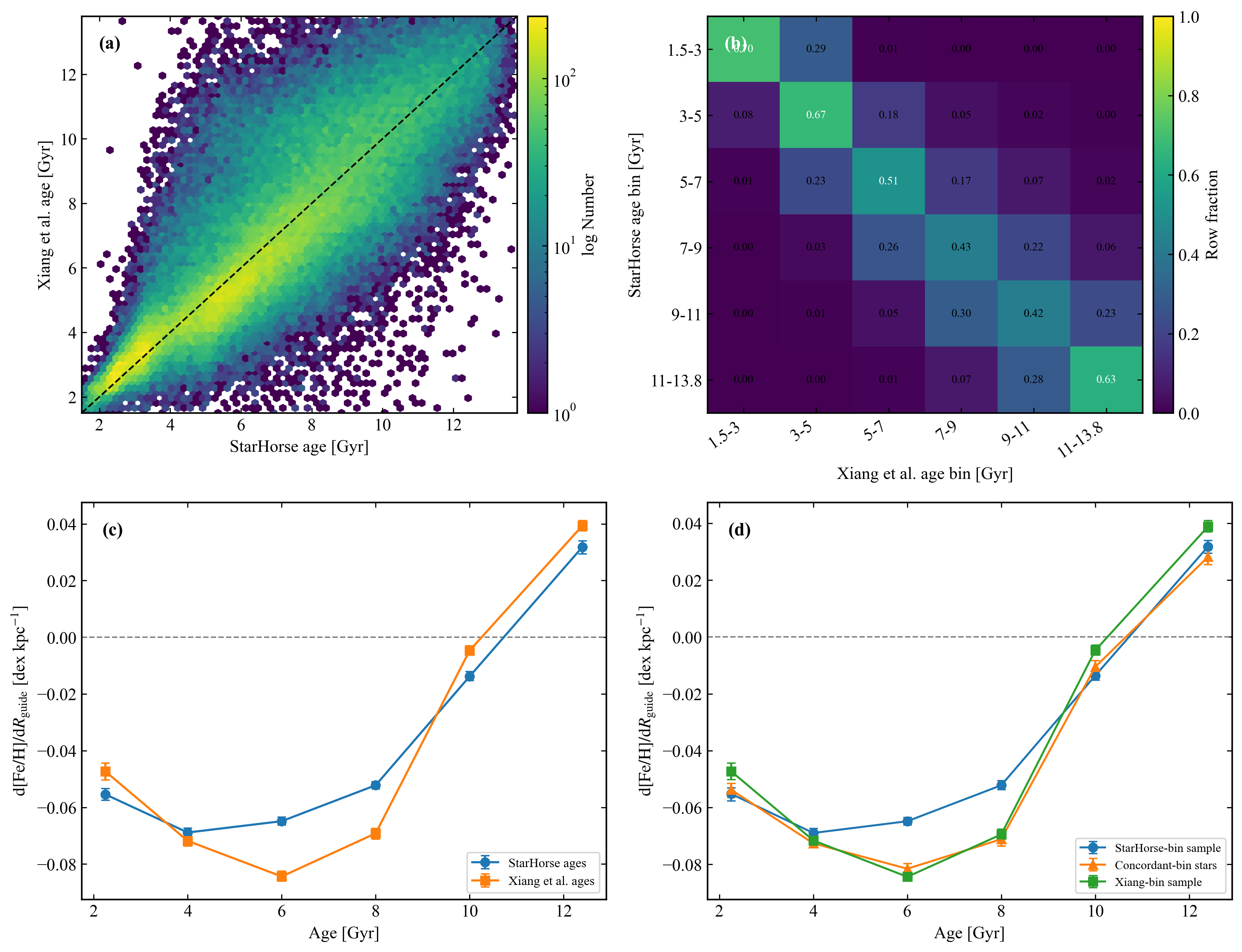}
	\caption{\rev{Sensitivity of the MGAR to the adopted stellar-age scale. Panel (a) compares the StarHorse \citep{Queiroz2023} with the Xiang et al. \citep{Xiang2022,Xiang2025} ages for the common-star sample, and panel (b) shows the corresponding age-bin transition matrix. Panel (c) compares the MGARs obtained with the two age assignments, while panel (d) additionally shows the result for stars remaining in the same age bin under both assignments. All gradients are measured over $5.0<R_{\rm guide}<10.5$ kpc using the same fiducial guiding-radius definition, LAMOST [Fe/H] measurements, and fitting procedure.}}
	\label{fig:AppendiB_age_scale_robustness}
\end{figure*}

\rev{As a complementary empirical test of the sensitivity to the adopted age scale, we repeated the MGAR analysis using the stellar ages of \citet{Xiang2022}, adopting the updated catalogue presented by \citet{Xiang2025}. We retained the same common stars, LAMOST [Fe/H] measurements, fiducial guiding radii, radial fitting interval of $5.0<R_{\rm guide}<10.5$ kpc, and gradient-fitting procedure, with the adopted stellar ages being the only quantity changed in the MGAR calculation. Fig.~\ref{fig:AppendiB_age_scale_robustness}(a) shows a strong overall correlation between the StarHorse and Xiang et al. ages (Spearman $r\simeq0.83$), although appreciable star-by-star scatter is present. This scatter leads to substantial redistribution among the adopted age intervals, as shown by the transition matrix in Fig.~\ref{fig:AppendiB_age_scale_robustness}(b). Approximately 53 per cent of the stars within the common age range remain in the same age interval under the two age assignments, with substantial redistribution occurring particularly across the intermediate-age bins.}

\rev{Despite the substantial redistribution of stars between age bins, the broad intermediate-to-old-age behaviour of the MGAR is recovered with the alternative age scale. As shown in Fig.~\ref{fig:AppendiB_age_scale_robustness}(c), both age assignments yield substantially negative radial [Fe/H] gradients at intermediate ages followed by pronounced flattening towards older populations. The quantitative gradient amplitudes nevertheless differ between the two age assignments, demonstrating that the detailed MGAR morphology is sensitive to the adopted age estimator. The intermediate-to-old-age flattening, rather than the slope of any individual age bin, is therefore the more stable feature of this comparison.}

\rev{The age intervals showing appreciable changes in the MGAR also exhibit substantial changes in bin membership. Fig.~\ref{fig:AppendiB_age_scale_robustness}(d) further shows that restricting the comparison to stars assigned to the same age interval by both estimates changes the quantitative gradients, indicating that differences in age-bin membership contribute to the quantitative differences between the two MGARs. Taken together, the marginal-age Monte Carlo experiment and the alternative-age comparison show that the principal intermediate-to-old-age flattening persists under the tested age uncertainties and the alternative age-scale choice, whereas the detailed gradient amplitudes remain sensitive to the adopted age estimator. Neither test constitutes a direct propagation of the full age--metallicity covariance.}

\section{Population-definition and stellar-parameter systematics}\label{Appendix:population_systematics}

The chemical composition of the stellar sample changes strongly with age, as shown in Section~\ref{SubSec:Age_Chemo_Dynamical}. Consequently, the observed MGAR may be affected not only by the method used to measure the radial gradient but also by systematic changes in the stellar populations entering different age bins. Section~\ref{SubSec:Radial_metallicity_gradient} showed that the intermediate-to-old-age flattening persists within a fiducially defined low-$\alpha$ population, whereas explicit control of the age-dependent $T_{\rm eff}$--$\log g$ distribution changes the gradient amplitudes and the detailed behaviour at younger ages. Here, we examine the sensitivity of these results to the chemical-population definition and to the atmospheric-parameter distribution in greater detail.

The main analysis uses fixed abundance boundaries, with stars satisfying $[\alpha/{\rm M}]<0.10$ assigned to the low-$\alpha$ population and those with $[\alpha/{\rm M}]>0.20$ assigned to the high-$\alpha$ population. These boundaries provide a simple and reproducible diagnostic separation, but the loci of the low- and high-$\alpha$ sequences vary with metallicity and therefore cannot be followed exactly by a single horizontal cut \citep{Hayden2017,Feuillet2019,Queiroz2023}. We therefore construct an alternative metallicity-dependent classification by fitting the $[\alpha/{\rm M}]$ distribution in narrow metallicity intervals with two-component Gaussian-mixture models.

\begin{figure*}
	\centering
	\includegraphics[width=0.95\textwidth]{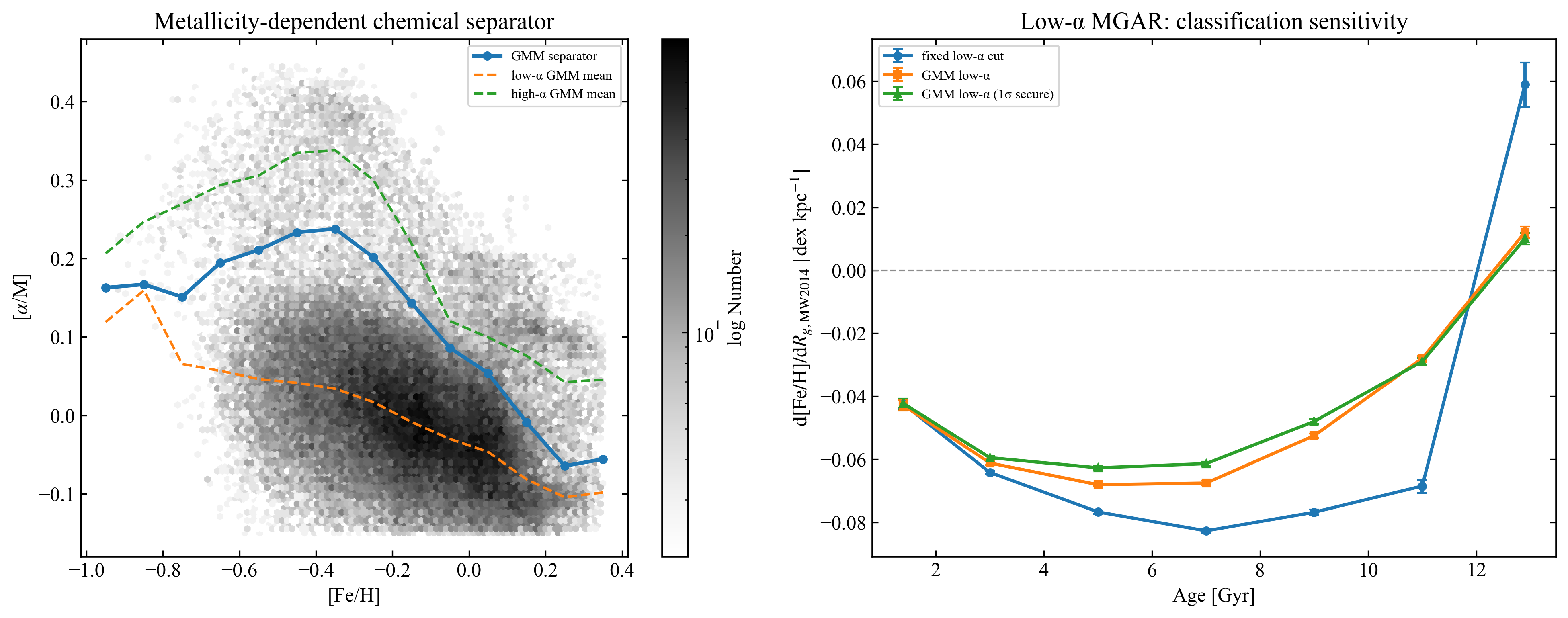}
	\caption{Sensitivity of the low-$\alpha$ MGAR to the chemical-population definition. The figure shows the metallicity-dependent Gaussian-mixture separator between the low- and high-$\alpha$ abundance sequences and compares the resulting low-$\alpha$ MGAR with those obtained using the fiducial fixed boundary and a more conservative Gaussian-mixture selection.}
	\label{fig:appendix_GMM_separator}
\end{figure*}

Fig.~\ref{fig:appendix_GMM_separator} shows the resulting metallicity-dependent separator and the corresponding low-$\alpha$ MGAR. The Gaussian-mixture classification follows the changing morphology of the abundance sequences more closely than the fixed abundance threshold. Although the numerical slopes differ between the classifications, the low-$\alpha$ population continues to exhibit relatively strong negative gradients at intermediate ages followed by substantial flattening towards older populations. A more conservative classification that excludes stars close to the mixture boundary yields the same qualitative age dependence. The intermediate-to-old-age flattening within the low-$\alpha$ population is therefore not produced by the particular fixed $[\alpha/{\rm M}]$ boundary adopted in the fiducial analysis.

\begin{figure}
	\centering
	\includegraphics[width=0.9\columnwidth]{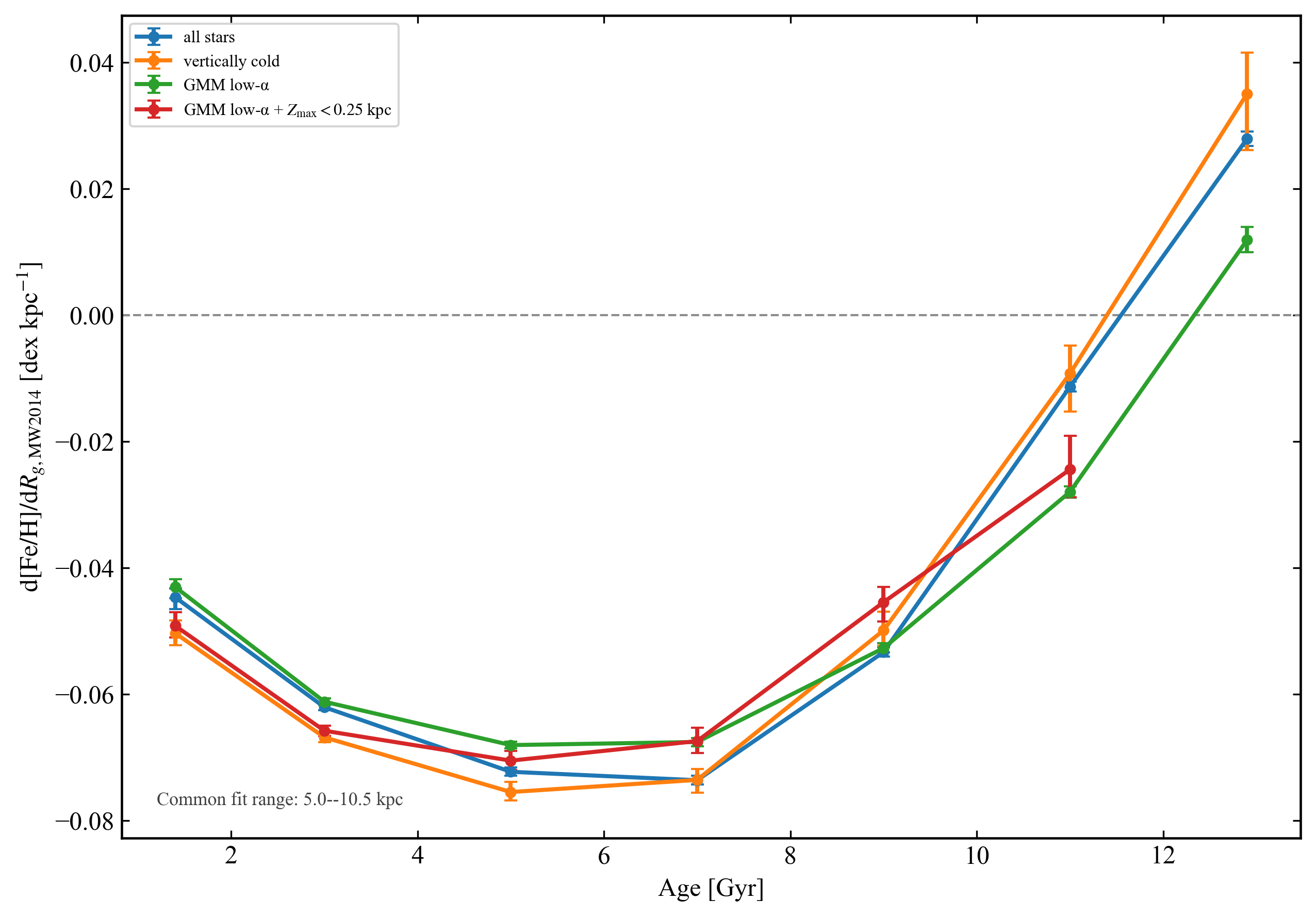}
	\caption{MGAR for the Gaussian-mixture low-$\alpha$ population after additionally requiring $Z_{\max}<0.25$ kpc and adopting the guiding radius calculated in MWPotential2014 \citep{Bovy2015}.}
	\label{fig:appendix_lowalpha_Zmax_MW}
\end{figure}

We further combine the Gaussian-mixture low-$\alpha$ classification with a vertically confined selection, $Z_{\max}<0.25$ kpc, and adopt the guiding radius calculated in MWPotential2014. This provides a restrictive test in which chemical-population mixing, vertical population mixing, and the use of the simplified $L_Z/V_{\rm circ}$ guiding-radius approximation are addressed simultaneously. The resulting MGAR is shown in Fig.~\ref{fig:appendix_lowalpha_Zmax_MW}. Despite the more restrictive selection and the associated reduction in sample size, relatively strong negative gradients remain at intermediate ages and are followed by substantial flattening towards older populations. The persistence of this behaviour shows that the intermediate-to-old-age flattening cannot be attributed solely to chemical-population mixing, vertical population mixing, or the simplified fiducial guiding-radius definition.

Because a two-component Gaussian mixture can formally be fitted even when the abundance distribution is only weakly bimodal, we also examine the statistical support for the adopted chemical decomposition. In each metallicity interval, the one- and two-component Gaussian descriptions are compared using the Bayesian information criterion (BIC), and the separation of the two fitted components is quantified by
\[
D=\frac{|\mu_1-\mu_2|}{\sqrt{(\sigma_1^2+\sigma_2^2)/2}},
\]
where $\mu_1$ and $\mu_2$ are the fitted component means and $\sigma_1$ and $\sigma_2$ are their dispersions. We define $\Delta{\rm BIC}={\rm BIC}_{1{\rm G}}-{\rm BIC}_{2{\rm G}}$, such that larger positive values indicate stronger statistical preference for the two-component model.

\begin{figure*}
	\centering
	\includegraphics[width=0.9\textwidth]{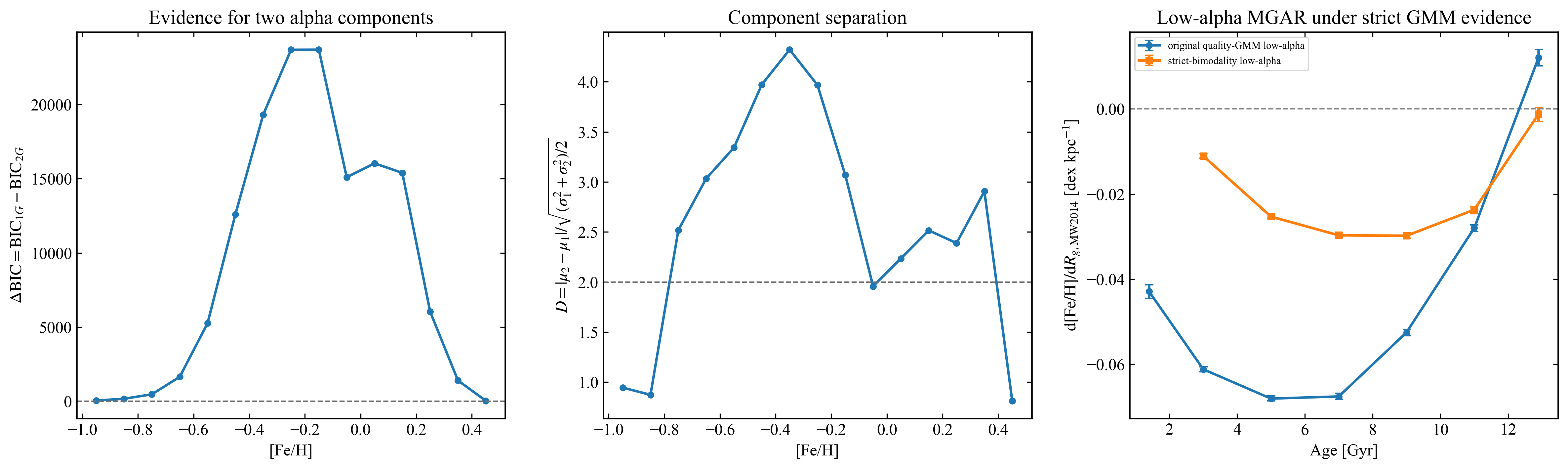}
	\caption{Statistical assessment of the two-component Gaussian-mixture description of the $[\alpha/{\rm M}]$ distribution. The figure shows the Bayesian-information-criterion difference, $\Delta{\rm BIC}={\rm BIC}_{1{\rm G}}-{\rm BIC}_{2{\rm G}}$, the separation statistic $D$ of the fitted abundance components, and the diagnostic MGAR obtained when the classification is restricted to metallicity intervals with strong evidence for bimodality. The latter is shown only as a diagnostic and is not used for quantitative interpretation of the MGAR.}
	\label{fig:appendix_GMM_bimodality}
\end{figure*}

As shown in Fig.~\ref{fig:appendix_GMM_bimodality}, the two-component description is strongly favoured over a substantial metallicity range, and the fitted abundance components are well separated over much of this regime. The strength of the bimodality nevertheless varies with metallicity. For completeness, the figure also shows the MGAR obtained after restricting the chemical classification to metallicity intervals with strong statistical evidence for two components. We do not use this ``strict-bimodality'' MGAR for quantitative interpretation because selecting stars according to metallicity intervals in which bimodality is detected directly truncates the [Fe/H] distribution, which is itself the dependent variable in the radial-gradient measurement. The BIC and component-separation diagnostics are therefore used to identify the metallicity range over which the Gaussian-mixture decomposition is statistically meaningful rather than to define an additional gradient sample.

\begin{figure*}
	\centering
	\includegraphics[width=0.9\textwidth]{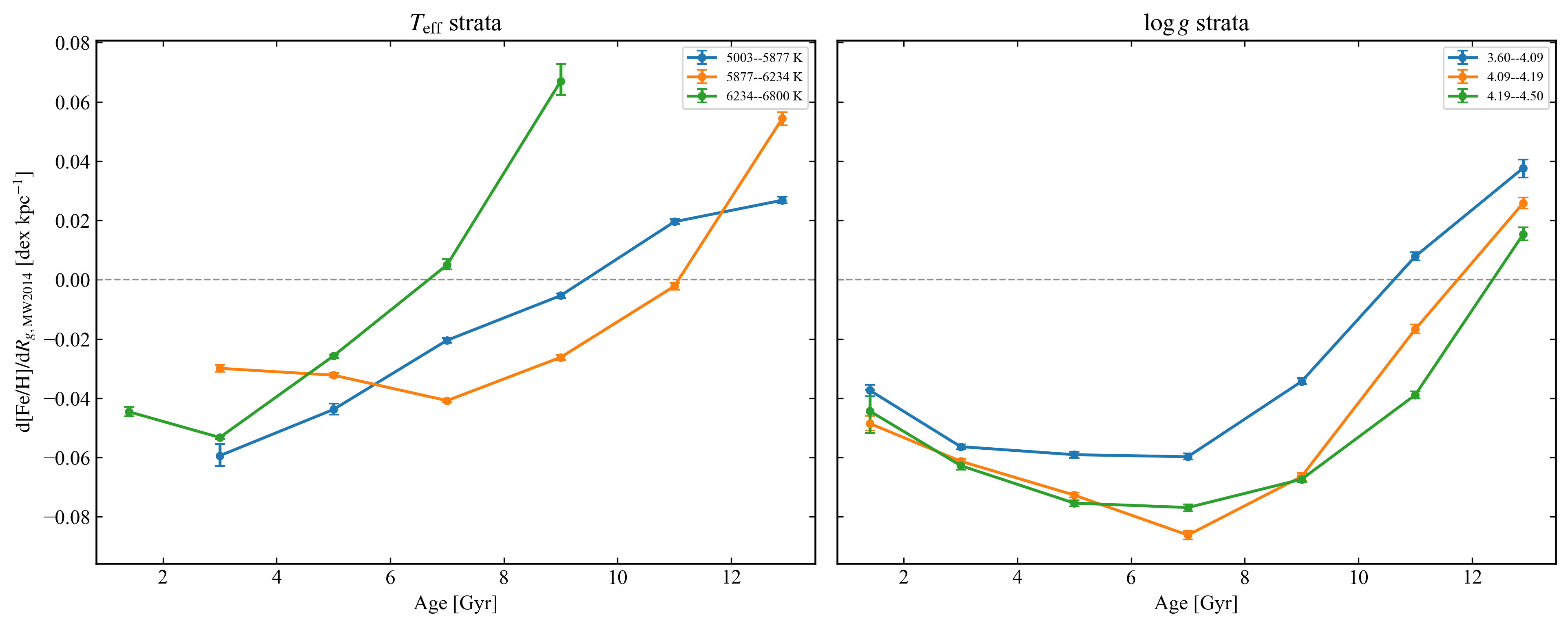}
	\caption{Dependence of the MGAR on stellar atmospheric parameters. The figure compares the MGAR measured in broad effective-temperature and surface-gravity intervals.}
	\label{fig:appendix_Teff_logg}
\end{figure*}

A separate population-related systematic arises because stellar age is coupled to atmospheric parameters in a dwarf- and turn-off-dominated sample. Different age intervals occupy systematically different regions of the $T_{\rm eff}$--$\log g$ plane, so part of the observed MGAR may reflect changes in the tracer population rather than age alone. To illustrate this coupling, Fig.~\ref{fig:appendix_Teff_logg} shows the MGAR measured independently in broad $T_{\rm eff}$ and $\log g$ intervals.

The effective-temperature subsamples show appreciable differences in the inferred gradient--age relation. The coolest stars exhibit comparatively shallow gradients over much of the sampled age range, whereas an intermediate-temperature subsample shows a more pronounced intermediate-age feature. The hottest stars are dominated by younger populations and provide limited coverage at older ages. These differences should not be interpreted as evidence that stars of different effective temperature intrinsically follow distinct Galactic chemical-evolution histories. Rather, $T_{\rm eff}$ is correlated with stellar mass, age, evolutionary state, and survey selection, so the temperature-stratified relations reveal a coupling between the stellar-parameter distribution and the inferred MGAR.

The corresponding $\log g$ subsamples show a more similar overall age dependence, although quantitative differences remain. These stratified tests demonstrate that broad atmospheric-parameter cuts do not produce equivalent tracer populations across stellar age. They therefore motivate the explicit two-dimensional control presented in Section~\ref{SubSec:Radial_metallicity_gradient}, in which the samples are first restricted to an overlapping $T_{\rm eff}$--$\log g$ domain and are then reweighted or hard matched. As shown there, controlling the $T_{\rm eff}$--$\log g$ distributions substantially reduces the absolute gradient amplitudes and modifies the detailed behaviour towards the younger end of the controlled interval, whereas the subsequent flattening towards older populations remains more stable.

\rev{Because spectroscopic abundance measurements can themselves exhibit systematic dependencies on stellar atmospheric parameters, we further assess the sensitivity of the controlled MGAR to parameter-dependent differences between the LAMOST and APOGEE metallicity scales. We cross-match the LAMOST sample with APOGEE DR17 \citep{Abdurrouf2022} and compare the LAMOST and APOGEE [Fe/H] measurements for the same stars. For this test, we retain the same stellar ages and MWPotential2014 guiding radii \citep{Bovy2015} and apply the same $T_{\rm eff}$--$\log g$ control to both metallicity measurements; as in the main analysis, this control substantially reduces, but does not eliminate, residual differences in the stellar-parameter distributions. This same-star comparison therefore provides a direct empirical test of the sensitivity of the controlled MGAR to the adopted spectroscopic metallicity scale while keeping the stellar sample, age assignments, orbital quantities, and parameter-control procedure fixed.}

\begin{figure*}
	\centering
	\includegraphics[width=0.9\textwidth]{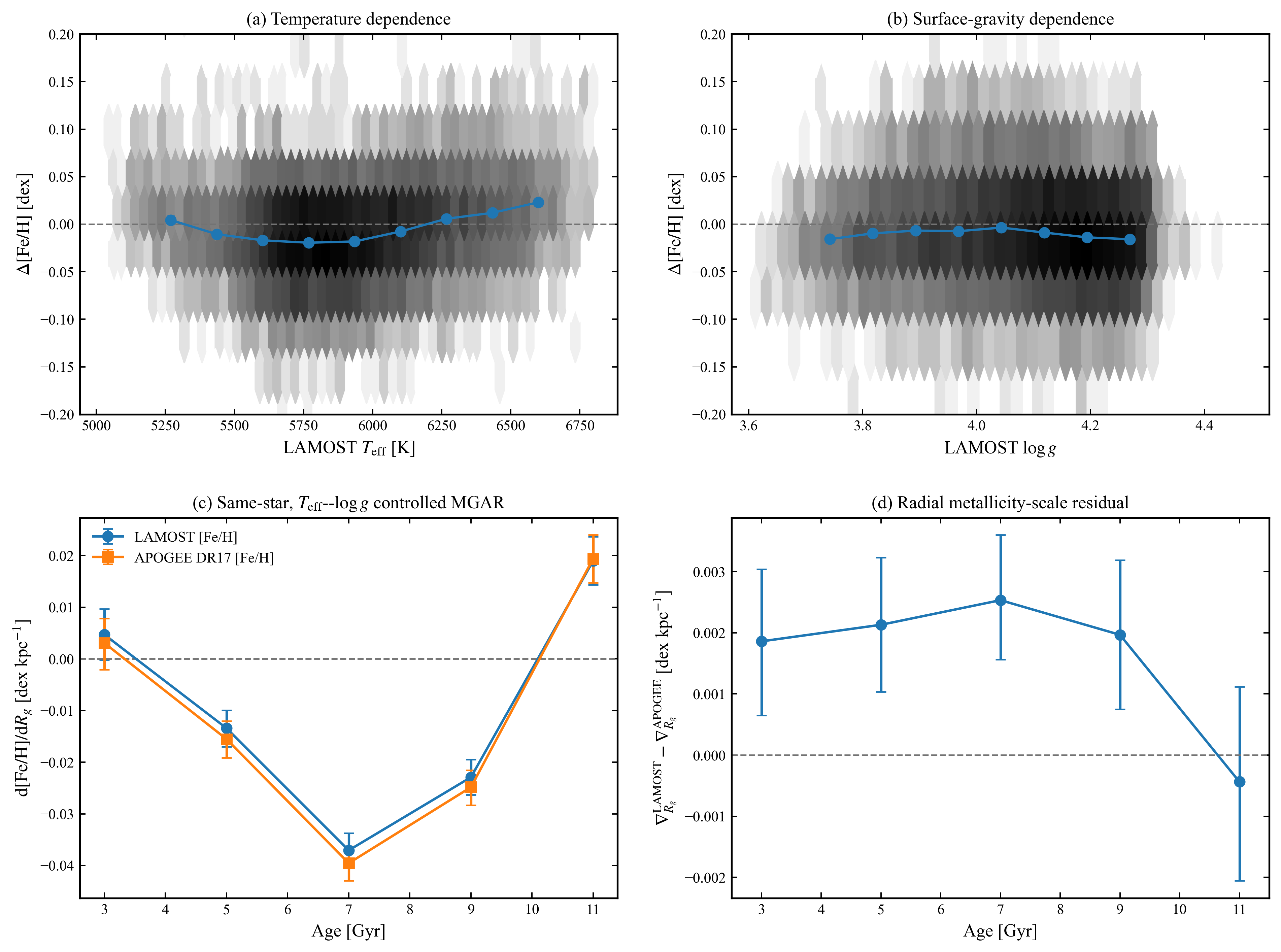}
    \caption{\rev{External test of the sensitivity of the controlled MGAR to the spectroscopic metallicity scale using stars in common between LAMOST and APOGEE DR17. Panels (a) and (b) show $\Delta[\mathrm{Fe/H}]=[\mathrm{Fe/H}]_{\rm LAMOST}-[\mathrm{Fe/H}]_{\rm APOGEE}$ as a function of $T_{\rm eff}$ and $\log g$, with blue symbols indicating the median trends. Panel (c) compares the $T_{\rm eff}$--$\log g$-controlled MGARs obtained using LAMOST and APOGEE metallicities for the same stars. Panel (d) shows the corresponding paired gradient difference, $\nabla_{R_g}^{\rm LAMOST}-\nabla_{R_g}^{\rm APOGEE}$, with 16th--84th percentile intervals from paired bootstrap resampling.}}
	\label{fig:Reviewer2_Fig_external_metallicity}
\end{figure*}

\rev{Panels (a) and (b) of Fig.~\ref{fig:Reviewer2_Fig_external_metallicity} show mild systematic variations in the LAMOST--APOGEE metallicity difference with atmospheric parameters. For the cross-matched sample of 11\,302 stars, the median difference, defined as $\Delta[\mathrm{Fe/H}]=[\mathrm{Fe/H}]_{\rm LAMOST}-[\mathrm{Fe/H}]_{\rm APOGEE}$, is approximately $-0.011$ dex, with a standard deviation of $0.066$ dex. The median offsets remain small over the bulk of the comparison sample, but their variation with $T_{\rm eff}$ and $\log g$ indicates that the two metallicity scales are not related by a strictly parameter-independent zero-point shift. This motivates comparing the controlled MGAR using the two abundance scales rather than assuming that parameter-dependent abundance systematics are negligible.}

\rev{The same-star comparison in Fig.~\ref{fig:Reviewer2_Fig_external_metallicity}(c) shows that replacing the LAMOST metallicities with APOGEE DR17 metallicities produces only small changes in the controlled MGAR. Both metallicity scales recover closely similar controlled MGARs over the comparison interval, including the subsequent flattening towards older populations. The paired gradient differences are of order $10^{-3}$ dex kpc$^{-1}$ and reach at most $\simeq2.5\times10^{-3}$ dex kpc$^{-1}$ from $3$ to $9$ Gyr, while the difference in the oldest controlled interval is consistent with zero [Fig.~\ref{fig:Reviewer2_Fig_external_metallicity}(d)]. Thus, although mild parameter-dependent offsets exist between the LAMOST and APOGEE metallicity scales, the broad age dependence of the controlled MGAR is recovered when APOGEE DR17 metallicities are adopted for the same stars. This indicates that the main features of the controlled MGAR are relatively insensitive to the adopted spectroscopic abundance scale, although the comparison does not directly establish whether the changes induced by the $T_{\rm eff}$--$\log g$ control are reproduced using APOGEE metallicities.}

These population-based tests clarify which aspects of the observed MGAR are sensitive to the definition of the stellar tracer. The persistence of the intermediate-to-old-age flattening under the fixed, Gaussian-mixture, conservative Gaussian-mixture, and vertically confined low-$\alpha$ selections shows that the full-sample relation cannot be explained solely by the changing mixture of chemically or vertically distinct populations. At the same time, the differences among the $T_{\rm eff}$-selected samples, together with the matched and reweighted analysis in the main text, show that the absolute gradient amplitudes and detailed behaviour at younger ages are sensitive to the stellar-parameter distribution. \rev{The same-star LAMOST--APOGEE comparison further shows that the broad age dependence of the controlled MGAR is recovered when APOGEE DR17 [Fe/H] measurements are adopted for the same stars. This indicates that the main features of the controlled MGAR are relatively insensitive to the adopted spectroscopic abundance scale, although residual abundance systematics and stellar-parameter selection effects cannot be excluded.} We therefore regard the intermediate-to-old-age flattening as more stable than the detailed young-age morphology or the absolute gradient amplitudes with respect to population-definition and stellar-parameter systematics.


\bsp	
\label{lastpage}
\end{document}